\documentclass[12pt]{article}
\usepackage{color}
\usepackage{amssymb,amsmath,comment}
\usepackage{graphicx}
\usepackage{braket}
\usepackage{epsfig}
\usepackage{here}
\usepackage{bm}
\graphicspath{{./Figures/}}
\newcommand{\ba}{\begin{align}}
\newcommand{\ea}{\end{align}}

\def\nn{\nonumber}

\def\bea{\begin{eqnarray}}
\def\eea{\end{eqnarray}}
 \makeatletter
\def\alt{\mathrel{\mathpalette\gl@align<}}
\def\agt{\mathrel{\mathpalette\gl@align>}}
\def\gl@align#1#2{\lower.6ex\vbox{\baselineskip\z@skip\lineskip\z@
\ialign{$\m@th#1\hfil##\hfil$\crcr#2\crcr\sim\crcr}}} \makeatother
\renewcommand{\thefootnote}{\fnsymbol{footnote}}
\begin{document}
\begin{flushright}
\end{flushright}
\vspace*{1.0cm}

\begin{center}
\baselineskip 20pt 
{\Large\bf 
Gravitational Waves from Phase Transition \\
in Minimal SUSY $U(1)_{B-L}$ Model
}
\vspace{1cm}

{\large 
Naoyuki Haba
\ and \ Toshifumi Yamada
} \vspace{.5cm}

{\baselineskip 20pt \it
Institute of Science and Engineering, Shimane University, Matsue 690-8504, Japan
}

\vspace{.5cm}

\vspace{1.5cm} {\bf Abstract} \end{center}

Many extensions of the Standard Model include a new $U(1)$ gauge group that is broken spontaneously at a scale much above TeV.
If a $U(1)$-breaking phase transition occurs at nucleation temperature of $O(100)$-$O(1000)$~TeV,
 it can generate stochastic gravitational waves in $O(10)$-$O(100)$ Hz range if $\beta_{\rm n}/H_{\rm n}=1000$, which
  can be detected by ground-based detectors.
Meanwhile, supersymmetry (SUSY) may play a crucial role in the dynamics of such high-scale $U(1)$ gauge symmetry breaking, because
 SUSY breaking scale is expected to be at TeV to solve the hierarchy problem.
In this paper, we study the phase transition of $U(1)$ gauge symmetry breaking in a SUSY model in the SUSY limit.
We consider a particular example, the minimal SUSY $U(1)_{B-L}$ model.
We derive the finite temperature effective potential of the model in the SUSY limit, study a $U(1)_{B-L}$-breaking phase transition,
 and estimate gravitational waves generated from it.

\thispagestyle{empty}

%\bigskip
\newpage
\renewcommand{\thefootnote}{\arabic{footnote}}
%\addtocounter{page}{-1}
\setcounter{footnote}{0}
%%%%%%%%%%%%%%%%%%%%%%%%%%
%\baselineskip 36pt
% Main body
%%%%%%%%%%%%%%%%%%%%%%%%%%
\baselineskip 18pt
%%%%%%%%%%%%%%%%%%%%%%%%%%

\section{Introduction}

Many extensions of the Standard Model (SM) include a new $U(1)$ gauge group that is broken spontaneously,
 important examples being the minimal $U(1)_{B-L}$ model~\cite{Davidson:1978pm,Marshak:1979fm,Mohapatra:1980qe},
 the left-right symmetric model~\cite{Mohapatra:1974gc,Senjanovic:1975rk} and Pati-Salam model~\cite{Pati:1974yy}.
Usually, there is no theoretical reason to expect that the breaking scale of such $U(1)$ gauge group is at TeV scale.
If the breaking scale is beyond the reach of new gauge boson searches at colliders,
 observation of stochastic gravitational waves generated from a $U(1)$-breaking phase transition
 is the key to testing such models~\cite{Dev:2016feu}.
This is because the nucleation temperature of the phase transition
 is encoded by the peak position of gravitational wave spectrum,
 and ground-based detectors such as
 Advanced LIGO~\cite{TheLIGOScientific:2014jea}, Advanced Virgo~\cite{TheVirgo:2014hva}, KAGRA~\cite{Somiya:2011np},
planned Einstein Telescope~\cite{Punturo:2010zz,Hild:2010id} and planned Cosmic Explorer~\cite{Evans:2016mbw}
 cover the region of $10-100$~Hz, which corresponds to
 nucleation temperature of $O(100)-O(1000)$~TeV if the speed of phase transition over the Hubble rate is 1000.
Recent work on gravitational waves from the breaking of a new visible $U(1)$ gauge group that occurs separately from electroweak symmetry breaking
 includes \cite{Okada:2018xdh,Brdar:2018num,Croon:2018kqn,Hasegawa:2019amx,Brdar:2019fur}.

If the breaking of a $U(1)$ gauge group occurs at a scale much above TeV,
 supersymmetry (SUSY) may play a crucial role in its dynamics,
 since we expect SUSY breaking scale to be at TeV to stabilize the electroweak scale with respect to Planck scale.
In this paper, therefore, we study the phase transition 
 of a $U(1)$ gauge symmetry breaking in a SUSY model and gravitational waves generated from it.
We work in the SUSY limit, namely, we assume that the nucleation temperature is above the SUSY breaking scale so that
 soft SUSY breaking terms are negligible in the study of phase transition.
For concreteness, we focus on the minimal SUSY $U(1)_{B-L}$ model
\footnote{
In a different context, gravitational waves in a SUSY $U(1)_{B-L}$ model has been discussed in Ref.~\cite{Buchmuller:2013lra}.
}, 
which is by itself highly motivated because it can explain the origin of the seesaw scale,
 and if $B-L$ is broken by even charges, $R$-parity is derived and accounts for the stability of dark matter.
To simplify our analysis on $U(1)_{B-L}$-breaking phase transition, we assume $R$-symmetry of the model.
$R$-symmetry is well-motivated by itself because
 one can forbid $\mu H_u H_d$ term by $R$-symmetry thereby solving the $\mu$-problem.

Although we concentrate on the minimal SUSY $U(1)_{B-L}$ model, 
 our study is applicable to a wide class of $U(1)$-gauge-extended SUSY models that contain superfields with $U(1)$ charge $+a$ and $-a$
 and a gauge singlet $S$ to achieve the $U(1)$ breaking.
Remarkably, this $U(1)$ need not be visible, i.e. the SM fields need not be charged under it, for the study of gravitational waves.

We comment in passing that SUSY models are more predictive than non-SUSY models about high-scale $U(1)$-breaking phase transitions.
This is because in non-SUSY models where scalar field $\phi$ breaks an extra $U(1)$, no symmetry forbids the Higgs portal term
($H$ denotes the SM Higgs field),
\bea
\lambda_{\phi H} \, \phi^\dagger\phi H^\dagger H.
\eea
Suppose $\phi$ develops a large vacuum expectation value (VEV) much above the electroweak scale.
To achieve the correct electroweak symmetry breaking, one has two options;
one fine-tunes the portal coupling $\lambda_{\phi H}$ so that the emergent mass term $\lambda_{\phi H}|\langle\phi\rangle|^2\,H^\dagger H$ is negligible compared to genuine Higgs mass term $m_H^2\,H^\dagger H$;
or one assumes that the genuine Higgs mass term nearly cancels the emergent mass term.
In the latter case, the study of the $U(1)$-breaking phase transition involves the SM Higgs field and depends on unknown genuine Higgs mass term $m_H^2$, in addition to the $U(1)$-breaking scale.
This dependence on the genuine Higgs mass term, or equivalently the Higgs portal coupling,
 has been studied in Ref.~\cite{Dev:2019njv}.
In SUSY models, the Higgs portal coupling is forbidden at the renormalizable level and hence is justifiably neglected.

This paper is organized as follows.
In Section~2, we explain the minimal SUSY $U(1)_{B-L}$ model, and derive the finite temperature effective potential for
$U(1)_{B-L}$-breaking VEVs.
In Section~3, we numerically compute the $O(3)$-symmetric Euclidean action for a high-temperature $U(1)_{B-L}$-breaking phase transition,
 calculate quantities that determine gravitational wave spectrum,
 and estimate stochastic gravitational waves generated from a $U(1)_{B-L}$-breaking phase transition.
Section~4 summarizes the paper.
\\

\section{Finite Temperature Effective Potential in the Minimal SUSY $U(1)_{B-L}$ Model}

\subsection{Minimal SUSY $U(1)_{B-L}$ Model}

The minimal SUSY $U(1)_{B-L}$ model is defined as follows:
The gauge group is $SU(3)_C\times SU(2)_L\times U(1)_Y\times U(1)_{B-L}$.
The field content is that of the minimal SUSY Standard Model (MSSM) plus three isospin-singlet neutrinos $N_i^c \ (i=1,2,3)$ and $\Phi,\overline{\Phi},S$ with the following charge assignments.
\bea
N_i^c:({\bf 1},{\bf 1},0,1), \ \ \ 
\Phi:({\bf 1},{\bf 1},0,-2), \ \ \ \overline{\Phi}:({\bf 1},{\bf 1},0,2), \ \ \ S:({\bf 1},{\bf 1},0,0)
\eea
As usual, for the MSSM fields, the lepton doublets $L_i$ have $B-L=-1$, the lepton singlets $E_i^c$ have $+1$,
the quark doublets $Q_i$ have $\frac{1}{3}$, the quark singlets $U_i^c,D_i^c$ have $-\frac{1}{3}$, and the Higgs fields $H_u,H_d$ have 0.
The most general superpotential reads
\begin{align}
W&=W_{\rm MSSM}+(Y_D)_{ij}\,H_uL_iN^c_j+Y_{Mi}\,\Phi N^c_iN^c_i
\\
&+\lambda\, S(\overline{\Phi}\Phi-\frac{v^2}{2})+\frac{m}{2}S^2+\frac{\kappa}{3}S^3.
\label{super}
\end{align}
Here, mass term $\mu_\Phi \, \overline{\Phi}\Phi$ is absorbed by a redefinition of $S$ and $m,\kappa$.
$Y_D$ is the neutrino Dirac Yukawa coupling,
 and $Y_M$ is the coupling that generates Majorana mass for the right-handed neutrinos after $U(1)_{B-L}$ breaking.
By a phase redefinition, we take $\lambda,v^2,Y_{Mi}$ to be real positive without loss of generality.

From now on, we assume $|m|^2\ll v^2$ and $|\kappa| \ll 1$.
This limit is obtained when the model has $R$-symmetry, 
under which superfield $S$ has $R=+2$ and $\Phi,\overline{\Phi}$ have $R=0$, and the matter superfields have $R=+1$ and the Higgs superfields $H_u,H_d$ have $R=0$.
Assuming $R$-symmetry is advantageous for explaining the smallness of $\mu$ in $\mu H_uH_d$.
In the rest of the paper, we neglect $|m|^2,\kappa$ and work with the $R$-symmetric superpotential,
\begin{align}
W&=W_{\rm MSSM}|_{{\rm without \ }\mu{\rm \mathchar`-term}}+(Y_D)_{ij}\,H_uL_iN^c_j+(Y_M)_{ij}\,\Phi N^c_iN^c_j
\\
&+\lambda\, S(\overline{\Phi}\Phi-\frac{v^2}{2}).
\end{align}
As the mechanism for SUSY breaking (at zero temperature) is beyond the scope of this paper,
 we do not discuss soft SUSY breaking gaugino mass.
The tree-level scalar potential involving $\Phi,\overline{\Phi},S$ reads
\footnote{
By abuse of notation, we denote the scalar component by the same character as the superfield.}
\begin{align}
V&=\left|\lambda \,S\overline{\Phi}+Y_{Mi}\,N_i^cN_i^c\right|^2+\lambda^2\left|S\Phi\right|^2+\lambda^2\left|\overline{\Phi}\Phi-\frac{v^2}{2}\right|^2
\\
&+\frac{1}{2}g_{B-L}^2\left(-2|\Phi|^2+2|\Phi|^2-\sum_i(|N_i^c|^2-|L_i|^2+|E_i^c|^2+\frac{1}{3}|Q_i|^2
-\frac{1}{3}|U_i^c|^2-\frac{1}{3}|D_i^c|^2)\right)^2
\end{align}
\\

\subsection{Finite Temperature Effective Potential}

To compute the one-loop effective potential at zero and finite temperature, we need the field-dependent mass eigenvalues for bosonic and fermionic components.
When SUSY is preserved, bosonic and fermionic components have the same set of mass eigenvalues.
However, since SUSY is already broken at finite temperature, we must also consider SUSY-breaking configurations of VEVs, e.g., the case with $\langle\Phi\rangle\langle\overline{\Phi}\rangle\neq \frac{v^2}{2}$ giving $F$-term SUSY breaking,
and the case with $\langle\Phi\rangle\neq\langle\overline{\Phi}\rangle$ giving $D$-term SUSY breaking.
So, we study the mass eigenvalues of bosonic and fermionic components separately.

We use Landau gauge for $U(1)_{B-L}$ gauge theory.

Before deriving the field-dependent mass eigenvalues, we assume that the VEVs \textit{at any temperature} satisfy
\bea
\langle\Phi\rangle\langle\overline{\Phi}\rangle=({\rm real \ positive}), \ \ \ \ \ \ \langle S\rangle =0, \ \ \ \ \ \langle N_i^c\rangle=0.
\eea
Then, we take advantage of the $U(1)_{B-L}$ symmetry to set both $\langle\Phi\rangle,\langle\overline{\Phi}\rangle$ to be real positive,
 and rewrite these VEVs as
\bea
\langle\Phi\rangle\equiv \frac{1}{\sqrt{2}}h, \ \ \ \ \ \langle\overline{\Phi}\rangle\equiv \frac{1}{\sqrt{2}}\bar{h} \ \ \ \ \ \ \  (h>0, \ \bar{h}>0).
\eea
The rest of the section is devoted to the study on the potential for $h,\bar{h}$.

The $(h,\bar{h})$-dependent mass eigenvalues for bosonic components are given as follows:
We decompose the scalar components of $\Phi,\overline{\Phi}$ as
$\Phi=\frac{1}{\sqrt{2}}(h+\phi+i\,a)$, $\bar{\Phi}=\frac{1}{\sqrt{2}}(\bar{h}+\bar{\phi}+i\,\bar{a})$
where $\phi,\bar{\phi}$ represent CP-even components and $a,\bar{a}$ CP-odd components.
The $(h,\bar{h})$-dependent mass matrix for $\phi,\bar{\phi}$ is
\bea
\frac{1}{2}\begin{pmatrix} % or pmatrix or bmatrix or Bmatrix or ...
      \phi & \bar{\phi} \\
   \end{pmatrix}
   {\cal M}^2_{\phi\bar{\phi}}
         \begin{pmatrix} % or pmatrix or bmatrix or Bmatrix or ...
            \phi \\
            \bar{\phi} \\
         \end{pmatrix} \ \ \ {\rm with} \ \ \  {\cal M}^2_{\phi\bar{\phi}}=\begin{pmatrix} % or pmatrix or bmatrix or Bmatrix or ...
        2g_{B-L}^2(3h^2-\bar{h}^2)+\frac{1}{2}\lambda^2\bar{h}^2  &  (-4g_{B-L}^2+\lambda^2)h\bar{h}-\frac{1}{2}\lambda^2 v^2 \\
          & -2g_{B-L}^2(h^2-3\bar{h}^2)+\frac{1}{2}\lambda^2h^2 \\
      \end{pmatrix},
      \label{phiphi}\nn\\
 \eea
and that for $a,\bar{a}$ is
\bea
   \frac{1}{2}\begin{pmatrix} % or pmatrix or bmatrix or Bmatrix or ...
      a & \bar{a} \\
   \end{pmatrix}
   {\cal M}^2_{a\bar{a}}
            \begin{pmatrix} % or pmatrix or bmatrix or Bmatrix or ...
            a \\
            \bar{a} \\
         \end{pmatrix} \ \ \ {\rm with} \ \ \ {\cal M}^2_{a\bar{a}}=
      \begin{pmatrix} % or pmatrix or bmatrix or Bmatrix or ...
        2g_{B-L}^2(h^2-\bar{h}^2)+\frac{1}{2}\lambda^2\bar{h}^2  & \frac{1}{2}\lambda^2 v^2 \\
          & -2g_{B-L}^2(h^2-\bar{h}^2)+\frac{1}{2}\lambda^2h^2 \\
      \end{pmatrix},
      \label{aa}\nn\\
 \eea
 from which mass eigenvalues are obtained by diagonalization.
The $(h,\bar{h})$-dependent masses for $S$, $N_i^c$ and the MSSM fields are
\bea
{\cal M}^2_S |S|^2 + {\cal M}^2_{N_i^c} |N_i^c|^2 + {\cal M}^2_L |L_i|^2 + {\cal M}^2_E |E_i^c|^2
+{\cal M}^2_Q |Q_i|^2+{\cal M}^2_U |U_i^c|^2+{\cal M}^2_D |D_i^c|^2
\label{sandmatter}
\eea
with
\bea
&&{\cal M}^2_S=\frac{1}{2}\lambda^2(h^2+\bar{h}^2),
\\
&&{\cal M}^2_{N_i^c}=g_{B-L}^2(-h^2+\bar{h}^2)+\frac{1}{2}Y_{Mi}^2\,h^2
\\
&&{\cal M}^2_L=-g_{B-L}^2(-h^2+\bar{h}^2), \ \ {\cal M}^2_E=g_{B-L}^2(-h^2+\bar{h}^2), \ \ 
{\cal M}^2_Q=-\frac{1}{3}g_{B-L}^2(-h^2+\bar{h}^2),
\nn\\
&&{\cal M}^2_U={\cal M}^2_D=\frac{1}{3}g_{B-L}^2(-h^2+\bar{h}^2).
\eea
Note that the mass of the MSSM fields solely comes from $D$-term SUSY breaking.
The $(h,\bar{h})$-dependent mass term for the $U(1)_{B-L}$ gauge boson $X_\mu$ is
\bea
\frac{1}{2}{\cal M}^2_X \, X_\mu X^\mu \ \ \  {\rm with} \ \ \ {\cal M}_X^2=4g_{B-L}^2(h^2+\bar{h}^2).
\label{gauge}
\eea

The $(h,\bar{h})$-dependent mass eigenvalues of fermionic components are given as follows.
Let $\psi_\Phi,\psi_{\bar{\Phi}},\psi_S,\psi_{N_i^c}$ denote the fermionic part of $\Phi,\overline{\Phi},S,N_i^c$, respectively,
 and let $\widetilde{X}$ denote $U(1)_{B-L}$ gaugino.
The $(h,\bar{h})$-dependent Majorana mass matrix for fermionic components is given by
\bea
   \frac{1}{2}
   \begin{pmatrix} % or pmatrix or bmatrix or Bmatrix or ...
     \psi_\Phi & \psi_{\bar{\Phi}} & \psi_S & \psi_{N_i^c} & \widetilde{X}
   \end{pmatrix}
   {\cal M}_F
      \begin{pmatrix} % or pmatrix or bmatrix or Bmatrix or ...
         \psi_\Phi \\
         \psi_{\bar{\Phi}} \\
         \psi_S \\
         \psi_{N_i^c} \\
         \widetilde{X}
      \end{pmatrix} \ \ 
{\rm with} \ \ {\cal M}_F=   \begin{pmatrix} % or pmatrix or bmatrix or Bmatrix or ...
      0 & 0 & \frac{1}{\sqrt{2}}\lambda\,\bar{h} & 0 & 2g_{B-L}\, h\\
        & 0  & \frac{1}{\sqrt{2}}\lambda\, h & 0 & -2g_{B-L}\, \bar{h}\\
        &     & 0  & 0 & 0 \\
        &     &      & \frac{1}{\sqrt{2}}Y_{Mi}\,h & 0 \\
        &     &      &  0 & 0
   \end{pmatrix}.
   \nn\\
   \eea
The mass eigenvalues are obtained by diagonalizing ${\cal M}_F^\dagger {\cal M}_F$,
 and are given by $4g_{B-L}^2(h^2+\bar{h}^2)$, $4g_{B-L}^2(h^2+\bar{h}^2)$,
$\frac{1}{2}Y_{Mi}^2\,h^2$, $\frac{1}{2}\lambda^2(h^2+\bar{h}^2)$, $\frac{1}{2}\lambda^2(h^2+\bar{h}^2)$.

It is easy to verify that when $h=\bar{h}=v$ so that SUSY is preserved,
 non-zero mass eigenvalues of bosonic components obtained from Eqs.~(\ref{phiphi})-(\ref{gauge})
 coincide with those of fermionic components (with the correct counting of degrees of freedom).

Finally, the finite temperature effective potential~\cite{Dolan:1973qd} for $h,\bar{h}$ is obtained as
\begin{align}
V_{\rm eff}(h,\bar{h};\,\mu,T)&=\frac{1}{4}\lambda^2(h\bar{h}-v^2)^2+\frac{1}{2}g_{B-L}^2(h^2-\bar{h})^2
\label{tree}\\
&+\frac{1}{64\pi^2}\sum_j M_{Bj}^4\left(\log\frac{M_{Bj}^2}{\mu^2}-\frac{3}{2}\right)-\frac{1}{64\pi^2}\sum_j M_{Fj}^4\left(\log\frac{M_{Fj}^2}{\mu^2}-\frac{3}{2}\right)
\label{one-loop}\\
&+\frac{T^4}{2\pi^2}\sum_jJ_B(\frac{M_{Bj}^2}{T^2})-\frac{T^4}{2\pi^2}\sum_jJ_F(\frac{M_{Fj}^2}{T^2}).
\label{finite-t}
\end{align}
Here, Eq.~(\ref{tree}) represents the tree-level potential.
Eq.~(\ref{one-loop}) is the one-loop effective potential at zero-temperature, with $\mu$ being the renormalization scale in $\overline{\rm DR} $ scheme.
Eq.~(\ref{finite-t}) is the temperature-dependent part of the potential,
with $J_B(x^2)=\int_0^\infty{\rm d}y\,y^2\log(1-\exp[y^2+x^2])$ and $J_F(x^2)=\int_0^\infty{\rm d}y\,y^2\log(1+\exp[y^2+x^2])$.
$M_{Bj}^2$ denote the $(h,\bar{h})$-dependent mass eigenvalues for bosonic components,
 obtained by diagonalizing Eqs.~(\ref{phiphi}),(\ref{aa}) and from Eqs.~(\ref{sandmatter})-(\ref{gauge}),
 with no duplication for real scalars, 2 duplications for complex scalars and 3 duplications for $X_\mu$ gauge boson.
$M_{Fj}^2$ denote the $(h,\bar{h})$-dependent mass eigenvalues for fermionic components, which are
 $4g_{B-L}^2(h^2+\bar{h}^2)$, $4g_{B-L}^2(h^2+\bar{h}^2)$,
 $\frac{1}{2}Y_{Mi}^2\,h^2$, $\frac{1}{2}\lambda^2(h^2+\bar{h}^2)$, $\frac{1}{2}\lambda^2(h^2+\bar{h}^2)$,
 with 2 duplications for each.

At temperature near or above the critical temperature, daisy diagrams cause breakdown of perturbation theory.
This problem is remedied by replacing the tree-level masses of bosonic components $M_{Bj}^2$ in Eqs.~(\ref{phiphi})-(\ref{gauge})
with loop corrected ones.
We follow Ref.~\cite{Carrington:1991hz} and only include $T^2$-proportional part of the one-loop correction
\footnote{
This recipe does not provide a good approximation at low temperature, since the decoupling of particles in the loop is not included~\cite{Curtin:2016urg,Mazumdar:2018dfl}.
The correct recipe is to solve a self-consistency equation derived from the finite temperature effective potential.
Ref.~\cite{Curtin:2016urg} has confirmed the appropriateness of the partial dressing procedure~\cite{Boyd:1993tz}.
Unfortunately, this procedure has not yet been extended to a multi-field case, which is our case.
}, and make
 the following replacements for the mass of the scalar components of $\Phi,\overline{\Phi},S,N_i^c$ and MSSM fields:
 \footnote{
 ${\cal M}^2_{Q_{1,2}},{\cal M}^2_{U_{1,2}}$ represent the corrected masses for 1st and 2nd generation quark doublets and up-type singlets,
 and ${\cal M}^2_{Q_{1,2}},{\cal M}^2_{U_{1,2}}$ represent those for 3rd generation.
Their difference is a large thermal mass via the top quark Yukawa coupling $y_t$.
}
\begin{align}
{\cal M}^2_{\phi\bar{\phi}}&\to {\cal M}^2_{\phi\bar{\phi}}
+   \frac{3}{2}\frac{T^2}{24}\begin{pmatrix} % or pmatrix or bmatrix or Bmatrix or ...
      8g_{B-L}^2+2Y_{Mi}^2+4\lambda^2 & 0 \\
      0 &  8g_{B-L}^2+4\lambda^2 \\
   \end{pmatrix}
+ \frac{3}{2}T^2 g_{B-L}^2   \begin{pmatrix} % or pmatrix or bmatrix or Bmatrix or ...
      1 & 0 \\
       0& 1 \\
   \end{pmatrix}
   \label{phiphi-t}
 \\
 {\cal M}^2_{a\bar{a}}&\to {\cal M}^2_{a\bar{a}}
+   \frac{3}{2}\frac{T^2}{24}\begin{pmatrix} % or pmatrix or bmatrix or Bmatrix or ...
      8g_{B-L}^2+2Y_{Mi}^2+4\lambda^2 & 0 \\
      0 &  8g_{B-L}^2+4\lambda^2 \\
   \end{pmatrix}
+ \frac{3}{2}T^2 g_{B-L}^2   \begin{pmatrix} % or pmatrix or bmatrix or Bmatrix or ...
      1 & 0 \\
       0& 1 \\
   \end{pmatrix}
 \\
 {\cal M}^2_S &\to {\cal M}^2_S+\frac{3}{2}\frac{1}{6}T^2\lambda^2
 \\
 {\cal M}^2_{N_i^c} &\to {\cal M}^2_{N_i^c}+\frac{3}{2}\frac{T^2}{12}(g_{B-L}^2+2Y_{Mi}^2)+\frac{3}{2}\frac{1}{4}T^2g_{B-L}^2
 \\
 {\cal M}^2_L &\to {\cal M}^2_L +\frac{3}{2}\frac{T^2}{12}(g_{B-L}^2+\frac{1}{4}g_Y^2+\frac{3}{4}g^2)
 +\frac{3}{2}\frac{1}{4}T^2(g_{B-L}^2+\frac{1}{4}g_Y^2+\frac{3}{4}g^2)
 \\
  {\cal M}^2_E &\to {\cal M}^2_E +\frac{3}{2}\frac{T^2}{12}(g_{B-L}^2+g_Y^2)
 +\frac{3}{2}\frac{1}{4}T^2(g_{B-L}^2+g_Y^2)
 \\
  {\cal M}^2_{Q_{1,2}} &\to  {\cal M}^2_{Q_{1,2}} + \frac{3}{2}\frac{T^2}{12}(\frac{1}{9}g_{B-L}^2+\frac{1}{36}g_Y^2+\frac{3}{4}g^2+\frac{4}{3}g_s^2)+\frac{3}{2}\frac{1}{4}T^2(\frac{1}{9}g_{B-L}^2+\frac{1}{36}g_Y^2+\frac{3}{4}g^2+\frac{4}{3}g_s^2)
  \\
  {\cal M}^2_{Q_{3}} &\to  {\cal M}^2_{Q_3} + \frac{3}{2}\frac{T^2}{12}(\frac{1}{9}g_{B-L}^2+\frac{1}{36}g_Y^2+\frac{3}{4}g^2+\frac{4}{3}g_s^2+2y_t^2)+\frac{3}{2}\frac{1}{4}T^2(\frac{1}{9}g_{B-L}^2+\frac{1}{36}g_Y^2+\frac{3}{4}g^2+\frac{4}{3}g_s^2)
 \\
 {\cal M}^2_D &\to {\cal M}^2_D +  \frac{3}{2}\frac{T^2}{12}(\frac{1}{9}g_{B-L}^2+\frac{1}{9}g_Y^2+\frac{4}{3}g_s^2)+\frac{3}{2}\frac{1}{4}T^2(\frac{1}{9}g_{B-L}^2+\frac{1}{9}g_Y^2+\frac{4}{3}g_s^2)
 \\
  {\cal M}^2_{U_{1,2}} &\to {\cal M}^2_{U_{1,2}} +  \frac{3}{2}\frac{T^2}{12}(\frac{1}{9}g_{B-L}^2+\frac{4}{9}g_Y^2+\frac{4}{3}g_s^2)+\frac{3}{2}\frac{1}{4}T^2(\frac{1}{9}g_{B-L}^2+\frac{4}{9}g_Y^2+\frac{4}{3}g_s^2)
  \\
    {\cal M}^2_{U_3} &\to {\cal M}^2_{U_3} +  \frac{3}{2}\frac{T^2}{12}(\frac{1}{9}g_{B-L}^2+\frac{4}{9}g_Y^2+\frac{4}{3}g_s^2+2y_t^2)+\frac{3}{2}\frac{1}{4}T^2(\frac{1}{9}g_{B-L}^2+\frac{4}{9}g_Y^2+\frac{4}{3}g_s^2)
    \label{u3-t}
\end{align}
Here, the factor $\frac{3}{2}$ on the second and third terms on the right hand side reflects the fact that in SUSY theories,
a bosonic loop correction is always accompanied by a fermionic loop correction with the same coupling constant,
and that $T^2$-part of the fermionic one-loop correction to a boson mass is half the bosonic one-loop correction,
and hence their sum is $\frac{3}{2}$ times the bosonic one.
The bosonic part (i.e. part without factor $\frac{3}{2}$) of the second term
 comes from one-loop corrections via $D$-term and $F$-term quartic couplings,
 and that of the third term comes from one-loop corrections via gauge couplings.
For the longitudinal component of the $U(1)_{B-L}$ gauge boson, we replace its mass, $({\cal M}^2_X)^L$, as
\begin{align}
({\cal M}^2_X)^L &\to ({\cal M}^2_X)^L+\frac{3}{2}\cdot8g_{B-L}^2T^2,
\end{align}
 while the mass of the transverse component is unchanged.

In the rest of the paper, we use the finite temperature effective potential Eqs.~(\ref{tree})-(\ref{finite-t})
 with replacements Eqs.~(\ref{phiphi-t})-(\ref{u3-t}), to study the $U(1)_{B-L}$-breaking phase transition in the minimal SUSY $U(1)_{B-L}$ model.
 \\

\subsection{Behavior of the Finite Temperature Effective Potential}

We numerically evaluate the finite temperature effective potential $V_{\rm eff}(h,\bar{h};\,\mu,T)$ 
Eqs.~(\ref{tree})-(\ref{finite-t}) (with replacements Eqs.~(\ref{phiphi-t})-(\ref{u3-t})) for several benchmark parameter sets
and study its behavior.
The benchmarks we take are
\bea
(\lambda,~g_{B-L},~Y_{M3})&=&(0.01,~0.4,~1), \ \ \ (0.1,~0.4,~1), \ \ \ (0.01,~0.15,~1), \ \ \ (0.01,~0.4,~0.1)
\nn\\
\eea
 and we fix $Y_{M1}=Y_{M2}=0$ (only one right-handed neutrino has a large Majorana Yukawa coupling).
We take $\mu=v$, which does not generate a large logarithm because $v$ is the only mass scale in the model.
Then, the potential scales with $v^4$, and depends on $h,\bar{h}$ and temperature $T$ only through dimensionless quantities $h/v,~\bar{h}/v,~T/v$.

\begin{figure}[H]
  \begin{center}
    \includegraphics[width=55mm]{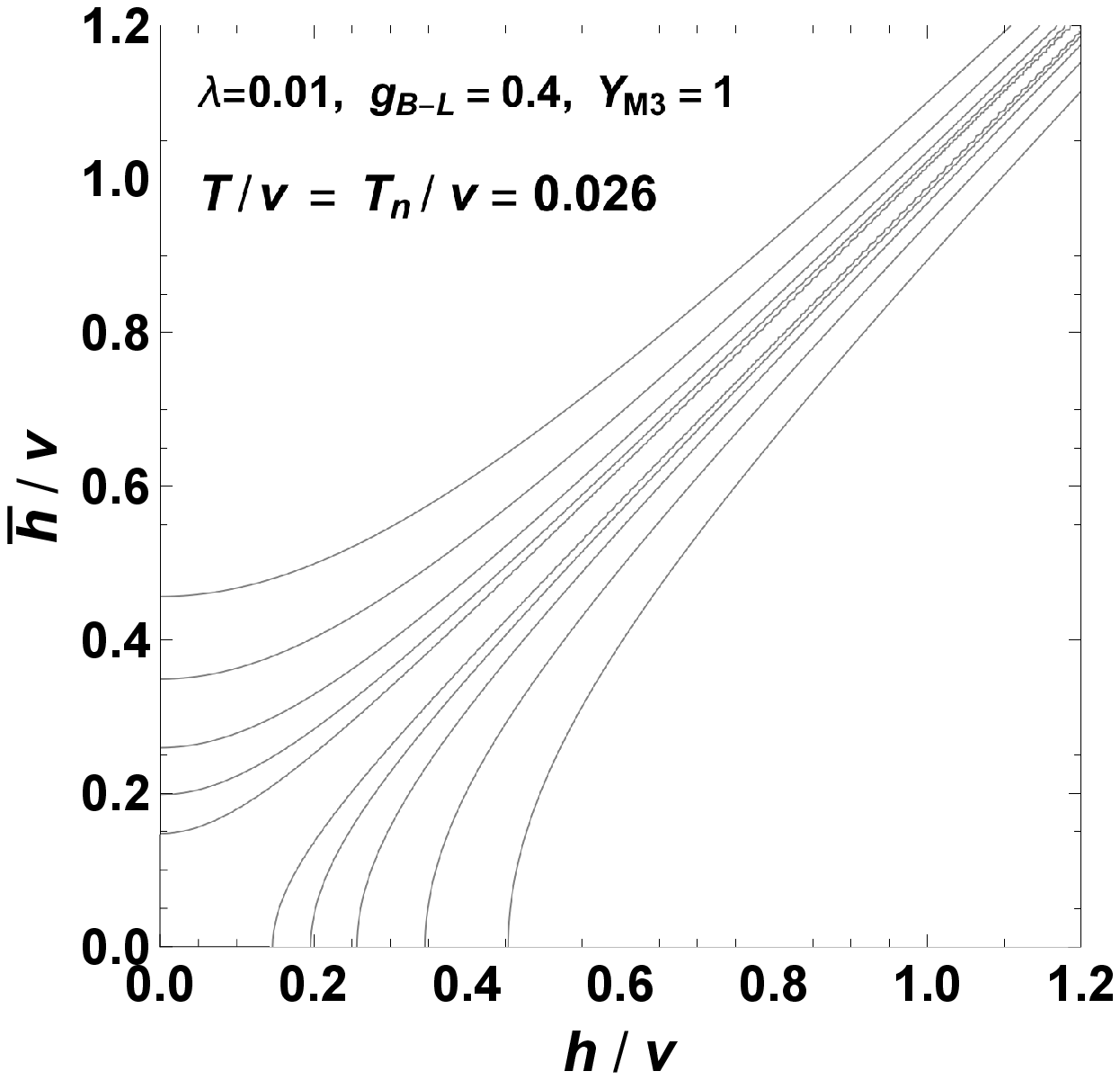}
    \includegraphics[width=55mm]{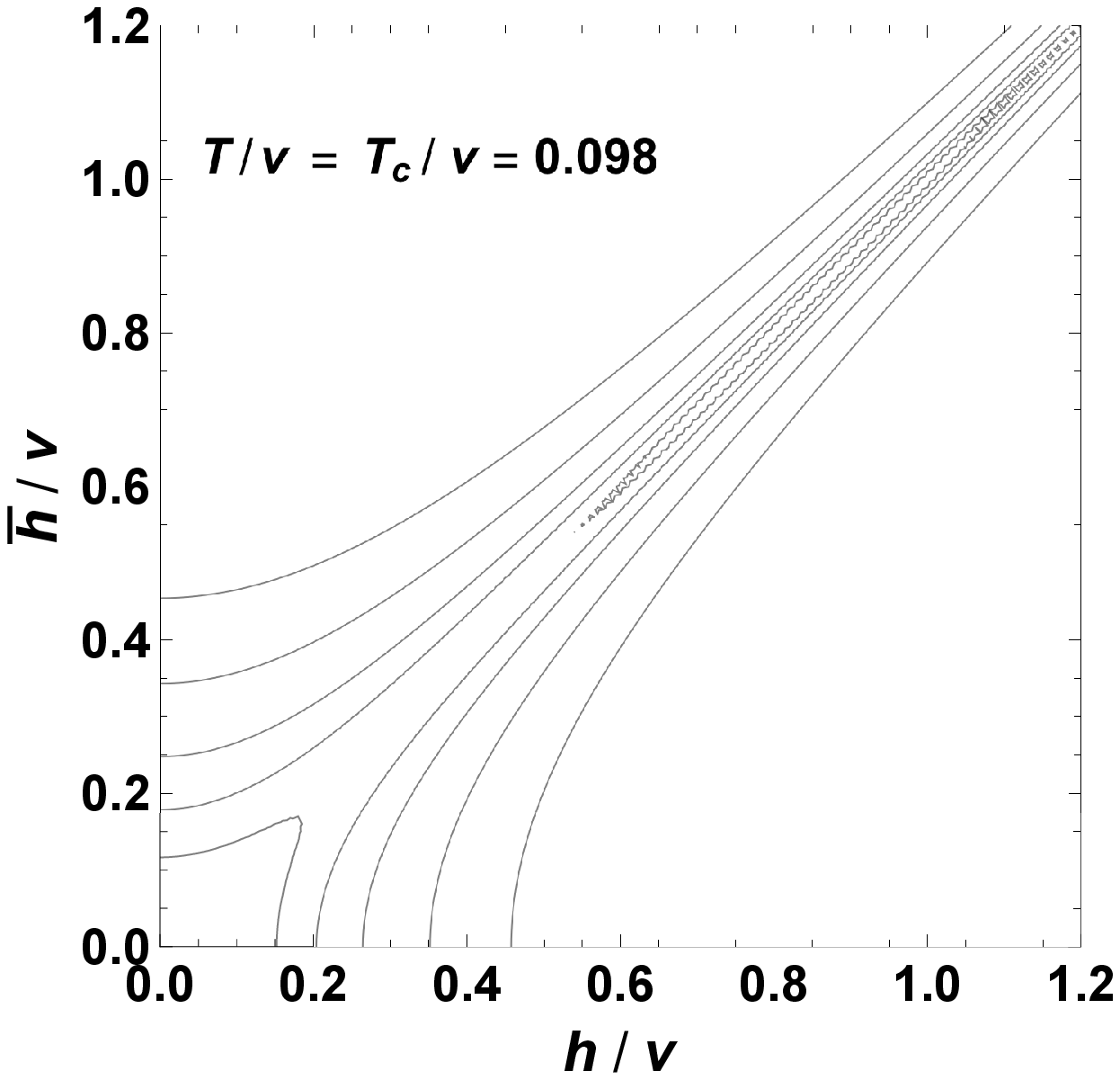}
    \includegraphics[width=55mm]{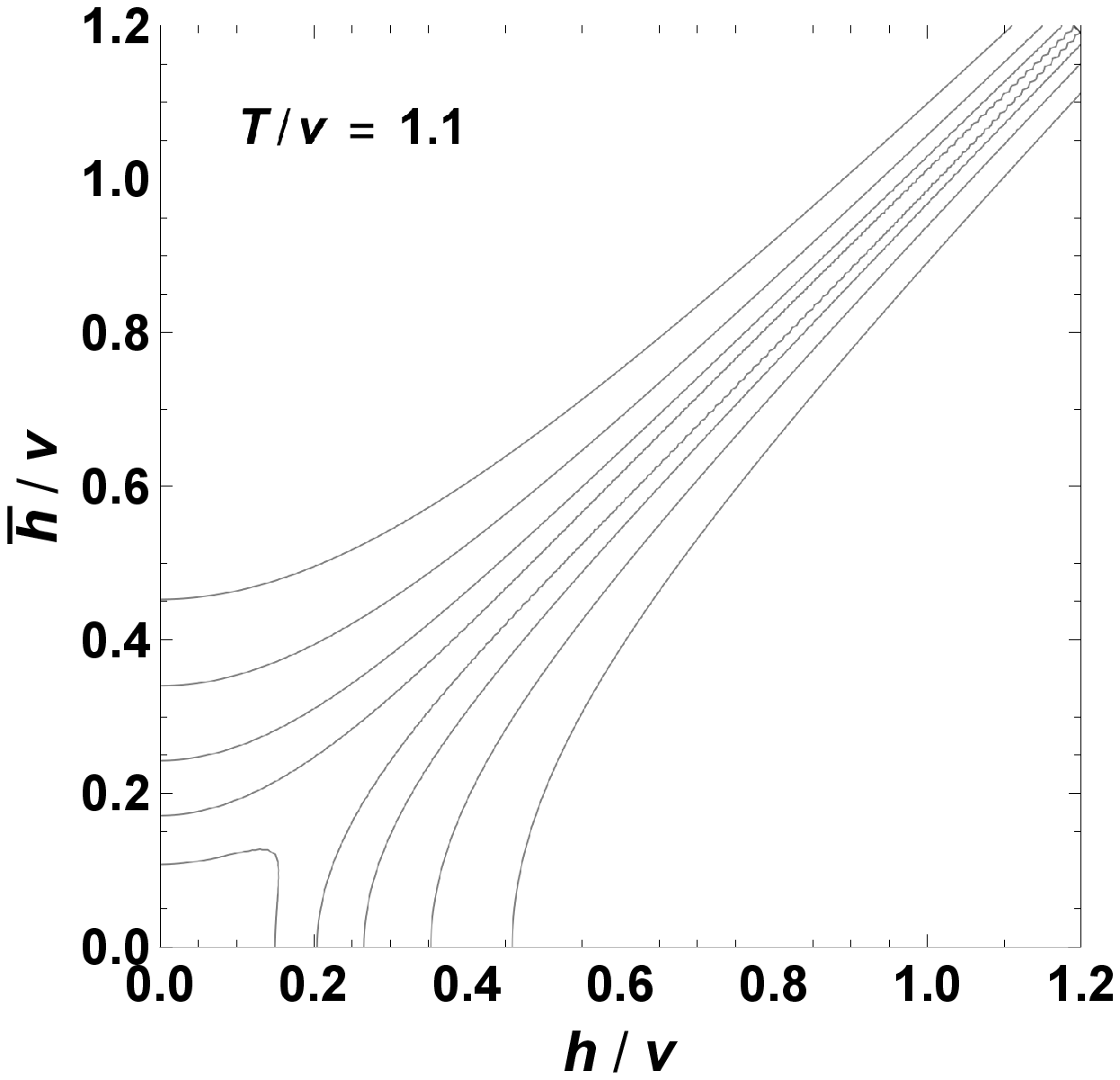}
    \caption{
    $V_{\rm eff}(h,\bar{h};\,\mu,T)-V_{\rm eff}(0,0;\,\mu,T)$ on the plane of $(h,\bar{h})$ at 
    the critical temperature $T=T_c=0.098 v$, at the nucleation temperature (which we will evaluate in Section~3) $T=T_{\rm n}=0.026 v$, and at a higher temperature $T=0.11v$,
    for the parameter set $(\lambda,~g_{B-L},~Y_{M3})=(0.01,~0.4,~1)$.
    The renormalization scale is set at $\mu=v$.
    The contours correspond, from outside to inside, to $V_{\rm eff}(h,\bar{h};\,\mu,T)-V_{\rm eff}(0,0;\,\mu,T)=
    3\cdot 10^{-3}v^4,\ 10^{-3}v^4,\ 3\cdot 10^{-4}v^4,\ 10^{-4}v^4,\ 3\cdot 10^{-5}v^4$.
    We caution that in the left panel, the barrier height is smaller than $3\cdot 10^{-5}v^4$ and hence is not visible.
    }
    \label{contour1}
  \end{center}
\end{figure}

\begin{figure}[H]
  \begin{center}
    \includegraphics[width=55mm]{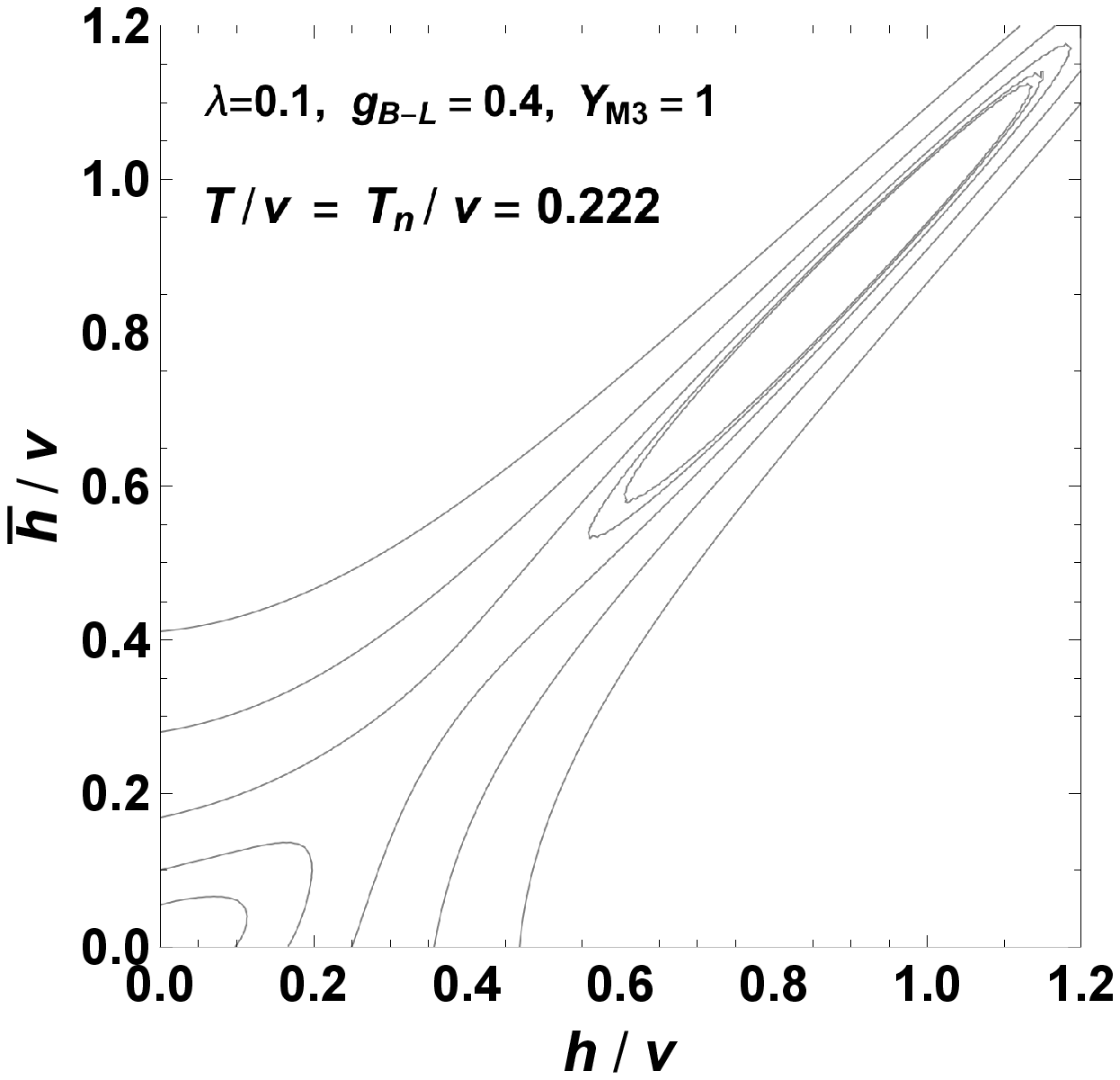}
    \includegraphics[width=55mm]{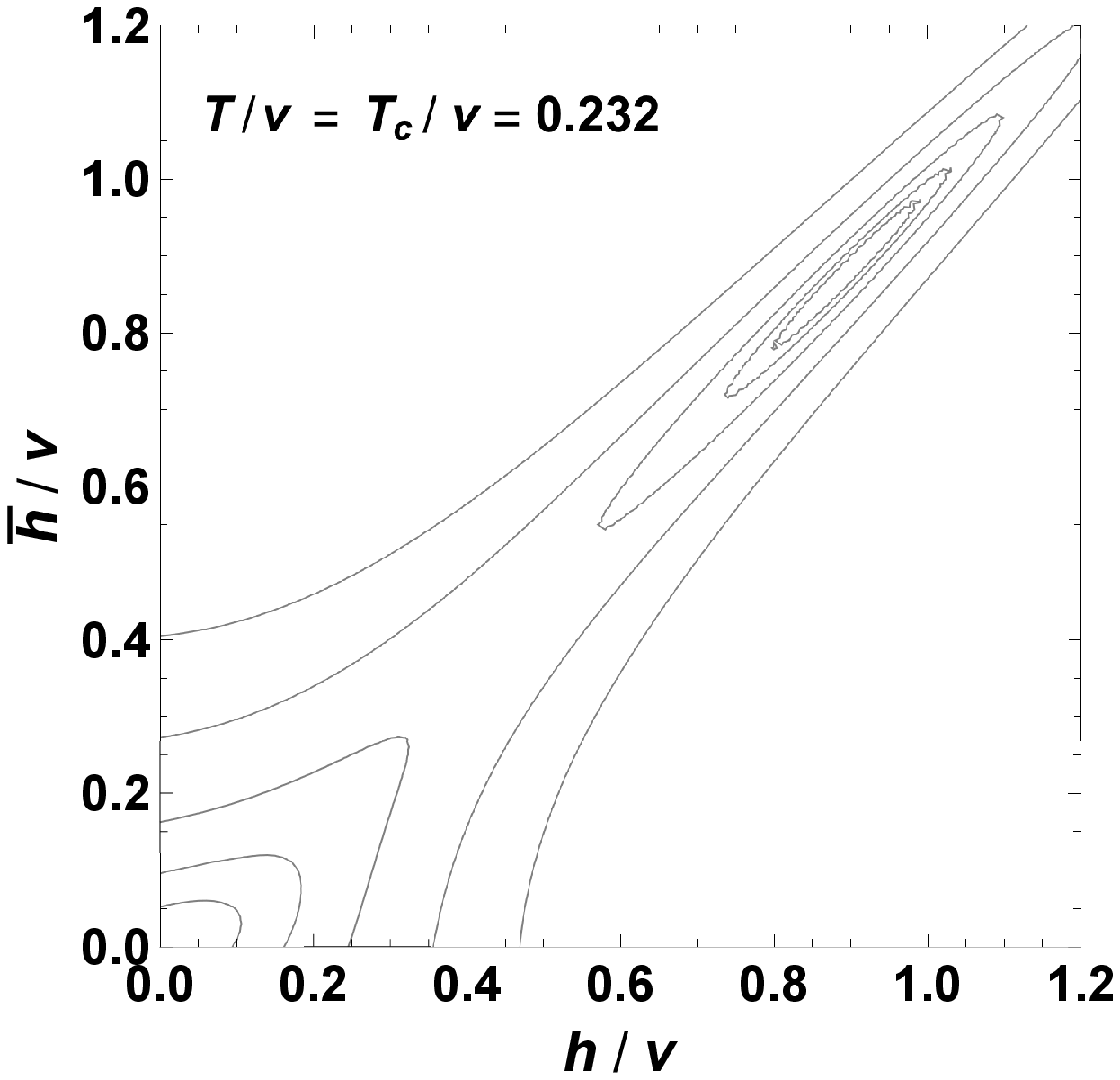}
    \includegraphics[width=55mm]{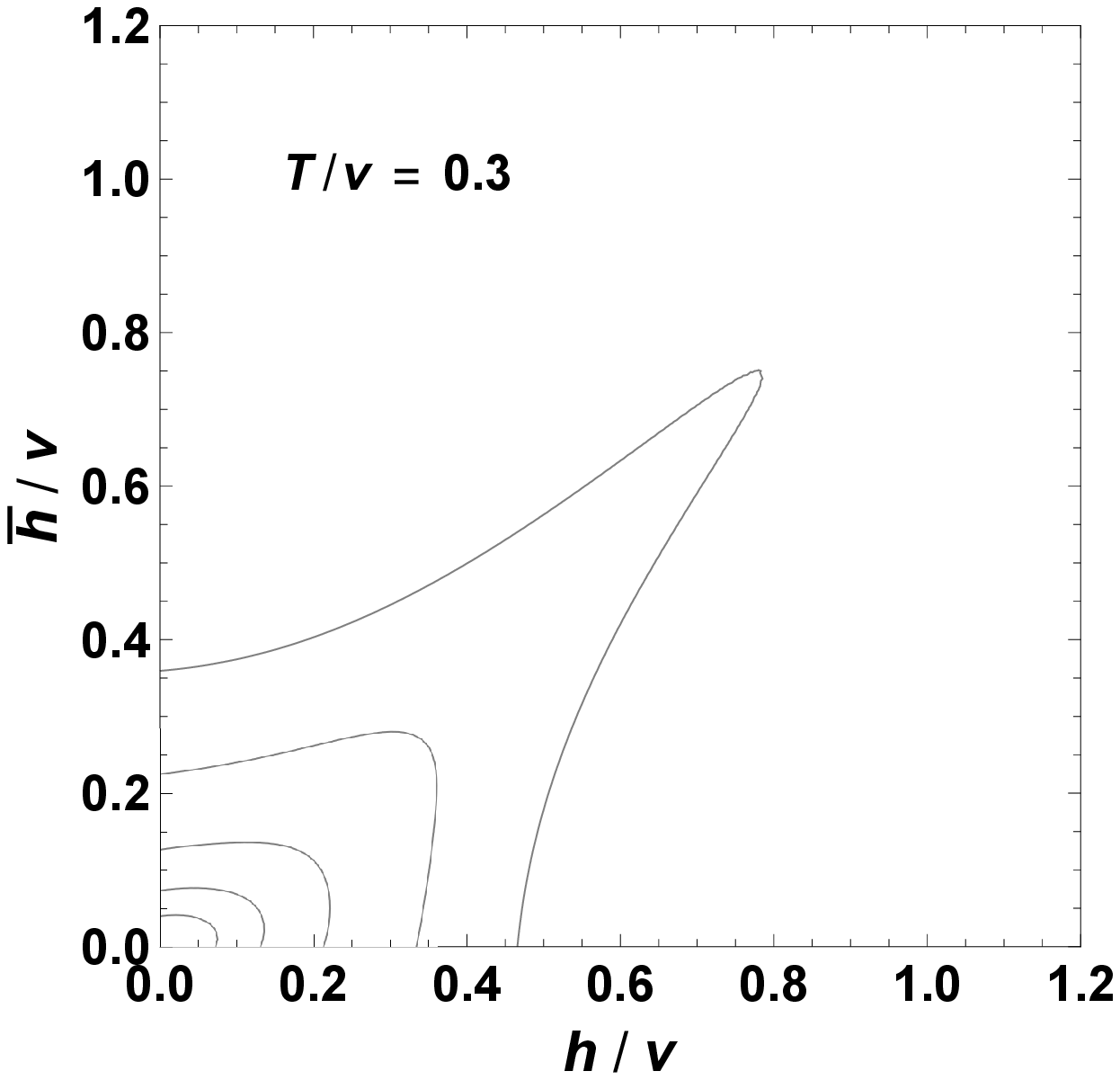}
    \caption{
    The same as Fig.~\ref{contour1}, except that
    the parameter set is $(\lambda,~g_{B-L},~Y_{M3})=(0.1,~0.4,~1)$, namely, $\lambda$ is ten times larger.
    The temperatures are taken at the critical temperature $T=T_c=0.232 v$, 
    the nucleation temperature $T=T_{\rm n}=0.222 v$, and a higher temperature $T=0.3 v$.
    The contours correspond, from outside to inside, to $V_{\rm eff}(h,\bar{h};\,\mu,T)-V_{\rm eff}(0,0;\,\mu,T)=
    3\cdot 10^{-3}v^4,\ 10^{-3}v^4,\ 3\cdot 10^{-4}v^4,\ 10^{-4}v^4,\ 3\cdot 10^{-5}v^4$.
    }
    \label{contour2}
  \end{center}
\end{figure}

\begin{figure}[H]
  \begin{center}
    \includegraphics[width=55mm]{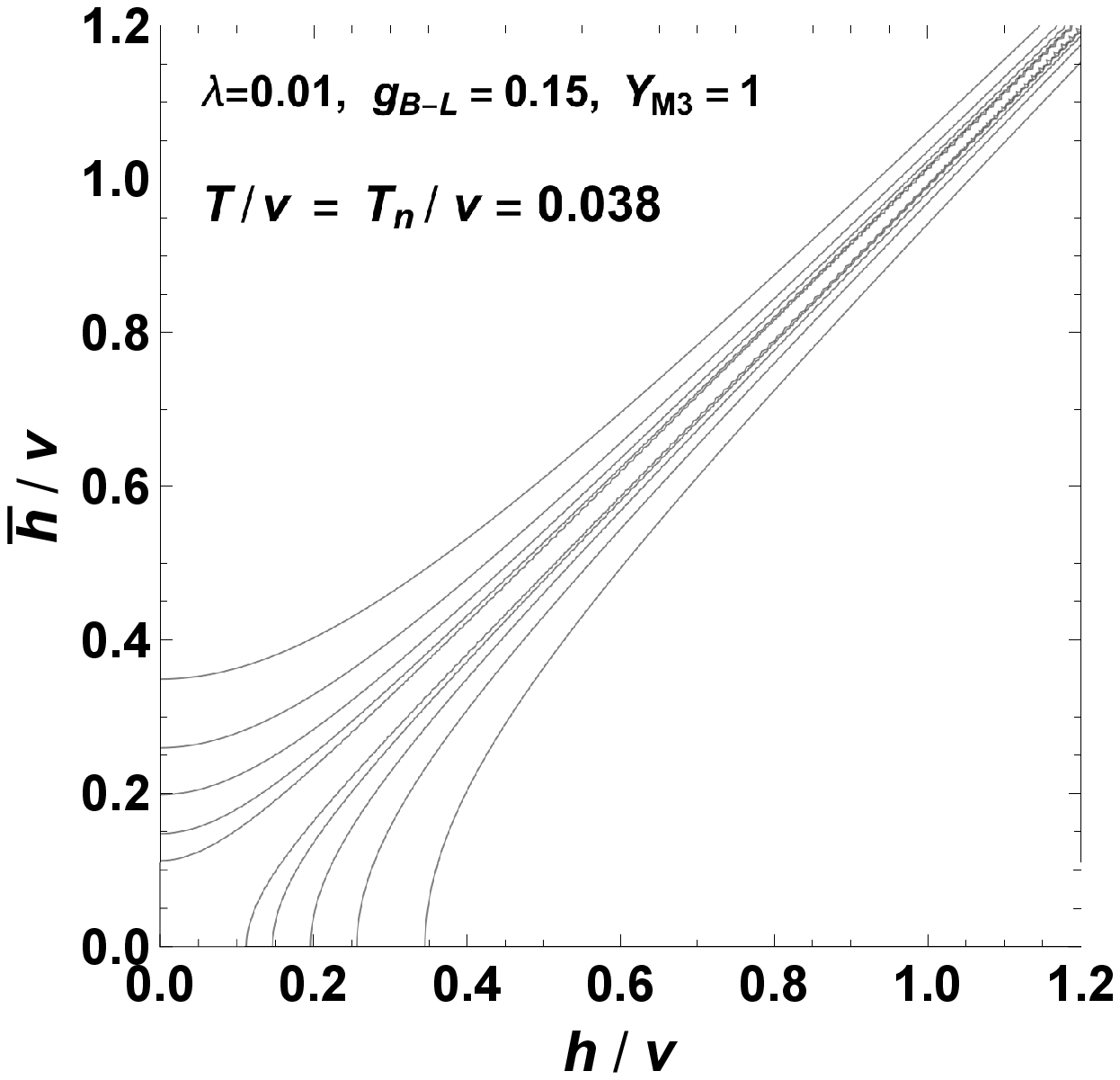}
    \includegraphics[width=55mm]{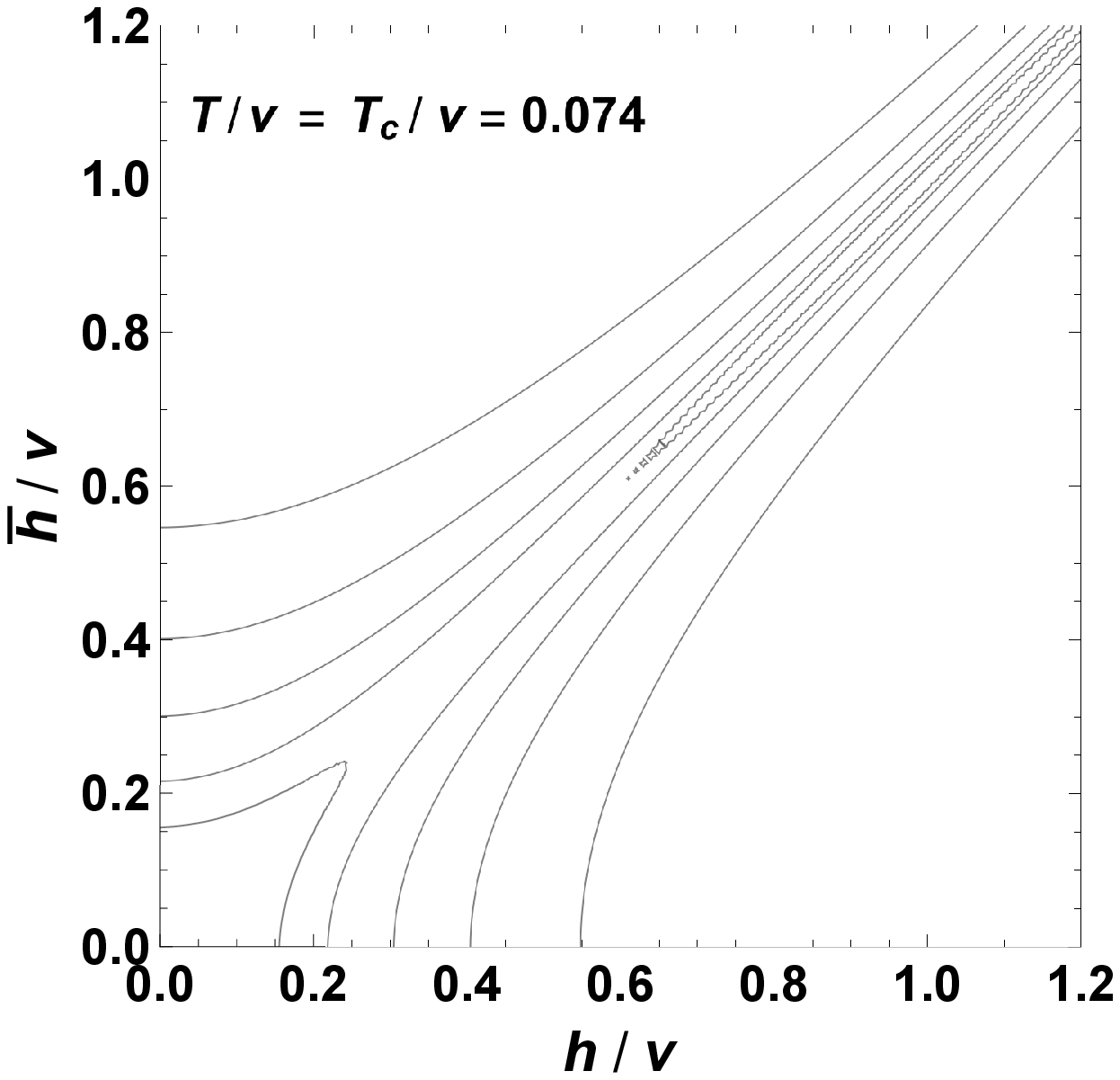}
    \includegraphics[width=55mm]{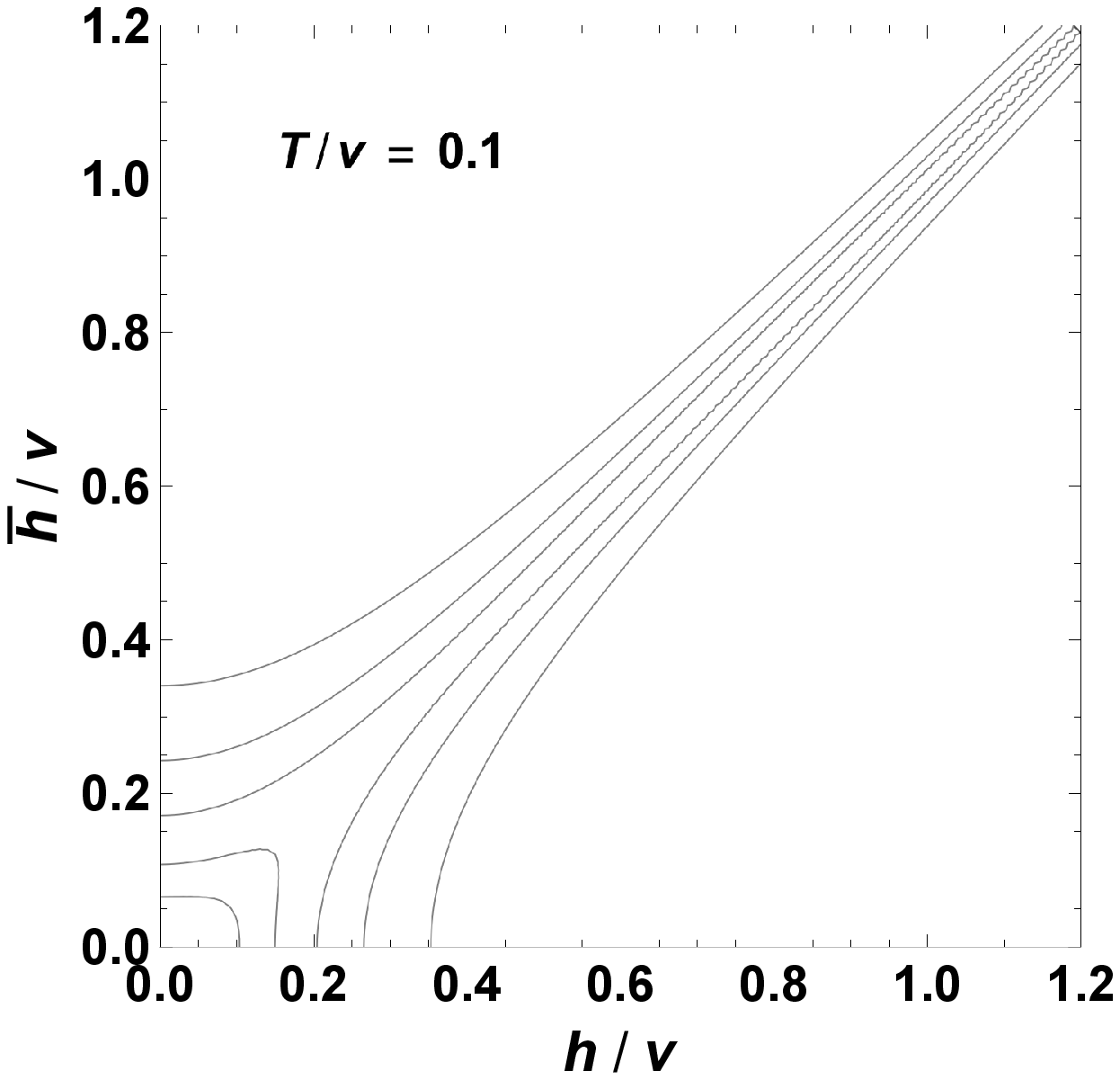}
    \caption{
    The same as Fig.~\ref{contour1}, except that
    the parameter set is $(\lambda,~g_{B-L},~Y_{M3})=(0.01,~0.15,~1)$, namely, $U(1)_{B-L}$ gauge coupling is smaller.
    The temperatures are taken at the critical temperature $T=T_c=0.074 v$, 
    the nucleation temperature $T=T_{\rm n}=0.038 v$, and a higher temperature $T=0.1 v$.
    The contours correspond, from outside to inside, to $V_{\rm eff}(h,\bar{h};\,\mu,T)-V_{\rm eff}(0,0;\,\mu,T)=
    10^{-3}v^4,\ 3\cdot 10^{-3}v^4,\ 10^{-3}v^4,\ 3\cdot 10^{-4}v^4,\ 10^{-4}v^4$.
    }
    \label{contour3}
  \end{center}
\end{figure}

\begin{figure}[H]
  \begin{center}
    \includegraphics[width=55mm]{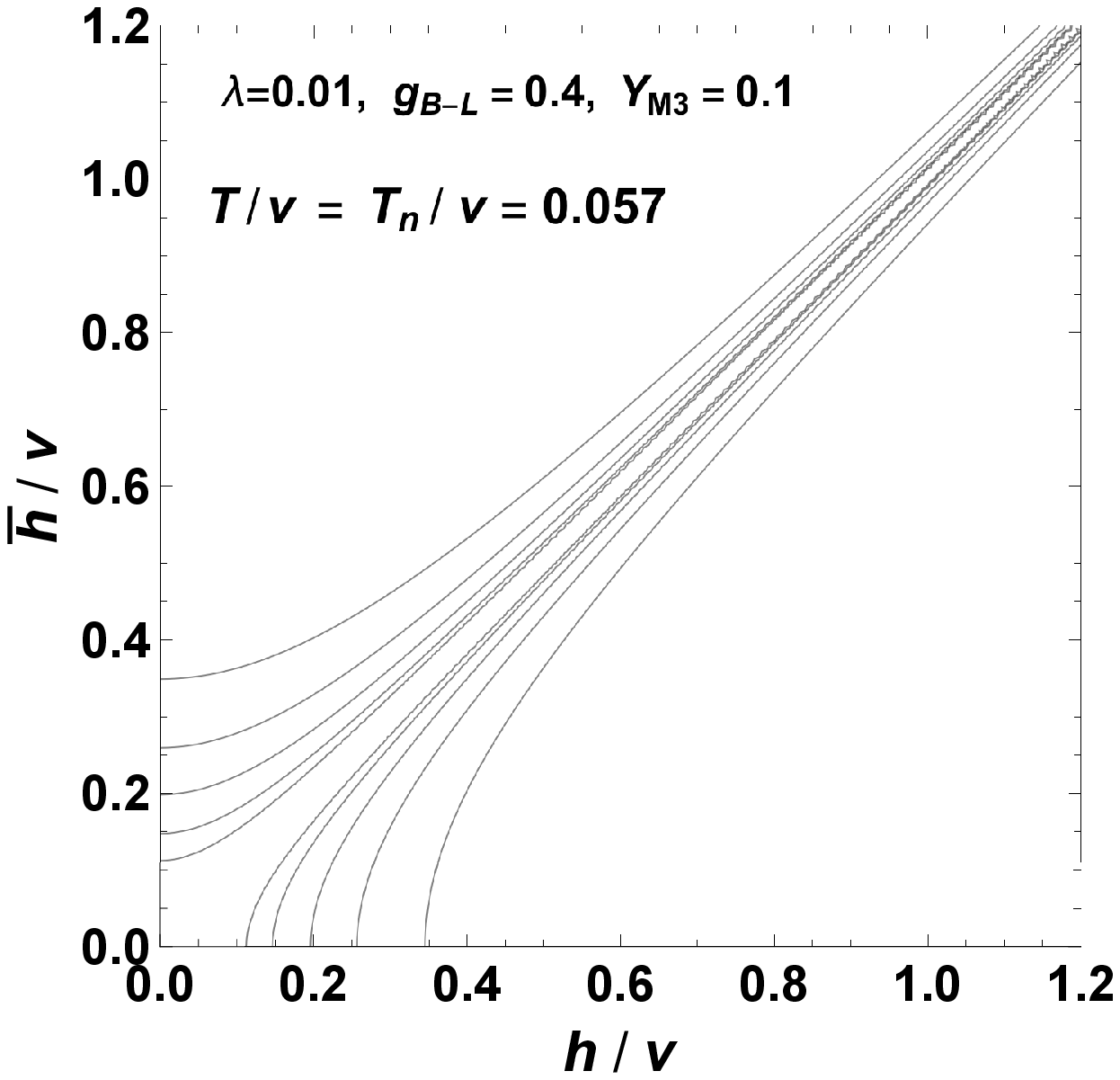}
    \includegraphics[width=55mm]{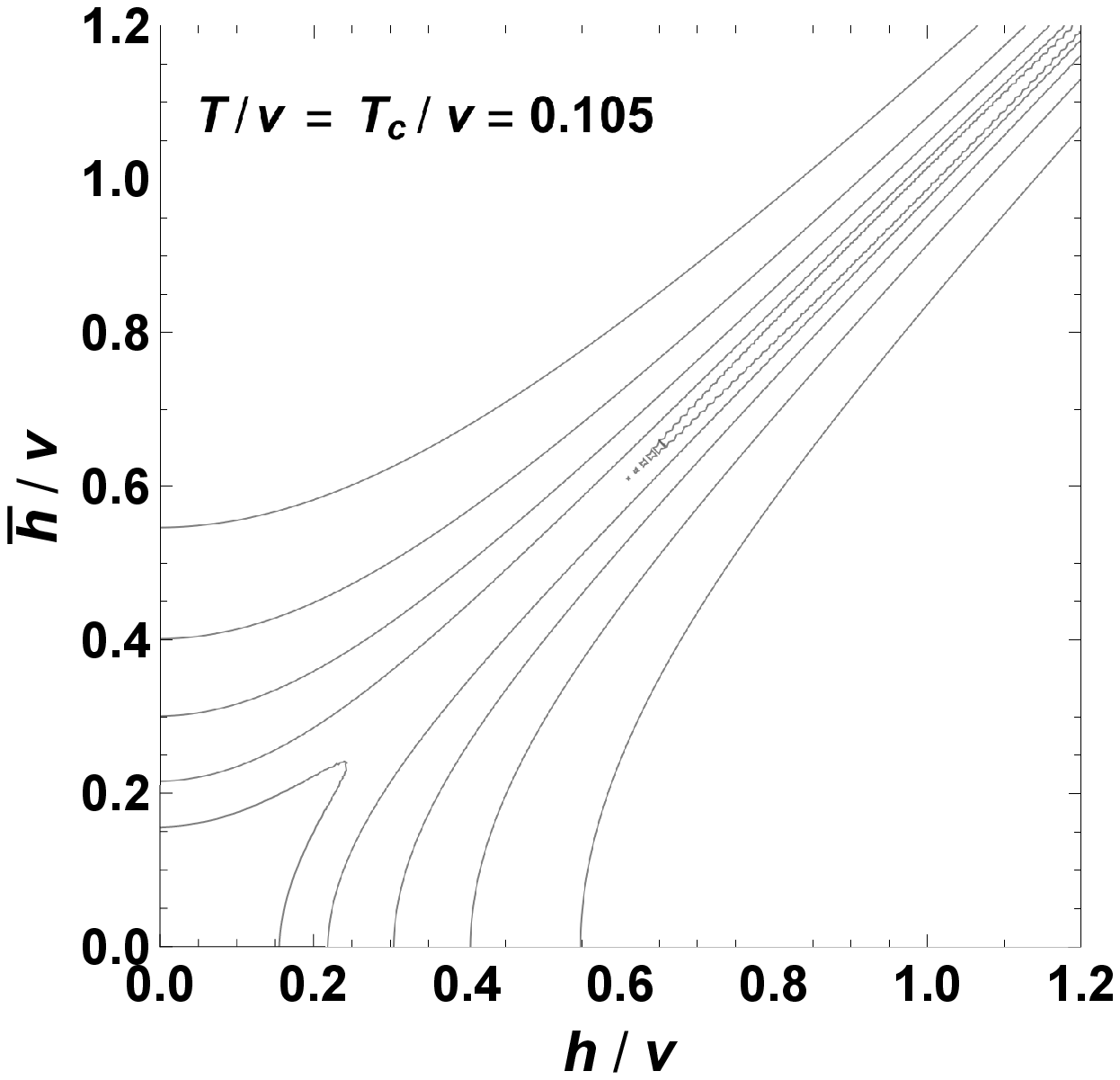}
    \includegraphics[width=55mm]{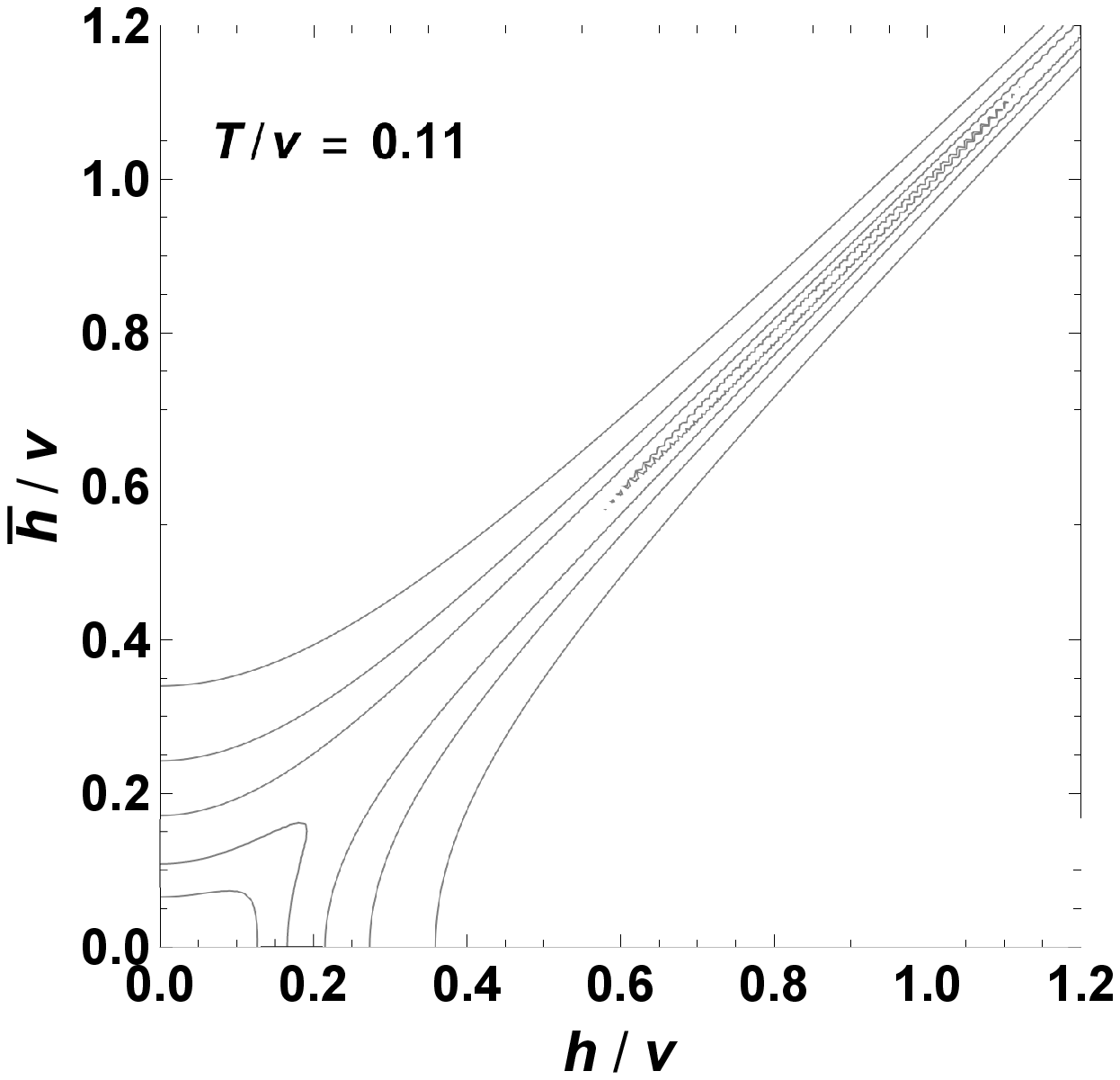}
    \caption{
    The same as Fig.~\ref{contour1}, except that
    the parameter set is $(\lambda,~g_{B-L},~Y_{M3})=(0.01,~0.4,~0.1)$, namely, Majorana Yukawa coupling $Y_{M3}$ is $1/10$ smaller.
    The temperatures are taken at the critical temperature $T=T_c=0.105 v$, 
    the nucleation temperature $T=T_{\rm n}=0.057 v$, and a higher temperature $T=0.11v$.
    $V_{\rm eff}(h,\bar{h};\,\mu,T)-V_{\rm eff}(0,0;\,\mu,T)=
    10^{-3}v^4,\ 3\cdot 10^{-3}v^4,\ 10^{-3}v^4,\ 3\cdot 10^{-4}v^4,\ 10^{-4}v^4$.
    }
    \label{contour4}
  \end{center}
\end{figure}

In Figs.~\ref{contour1},\ref{contour2},\ref{contour3},\ref{contour4}, 
we present $V_{\rm eff}(h,\bar{h};\,\mu,T)-V_{\rm eff}(0,0;\,\mu,T)$ on $(h,\bar{h})$ plane at the critical temperature $T=T_c$,
 at a high temperature slightly above $T_c$, and at the nucleation temperature $T_{\rm n}$ (which we will evaluate in the next section).

Figs.~\ref{contour1},\ref{contour2},\ref{contour3},\ref{contour4} show that the finite-temperature effective potential
  is nearly symmetric with respect to $h$ and $\bar{h}$ at temperature around or below the critical temperature $T_c$.
This indicates that even though only $\Phi$, not $\overline{\Phi}$, couples to the right-handed neutrino through Majorana Yukawa coupling $Y_{M3}$,
 this asymmetry does not affect the potential.

Since the potential is nearly symmetric with respect to $h$ and $\bar{h}$, we can approximate the classical tunneling path
 from the metastable vacuum $(h,\bar{h})=(0,0)$ to an absolute vacuum $(h,\bar{h})\neq(0,0)$ by the line $h=\bar{h}$,
 because $\partial_hV_{\rm eff}-\partial_{\bar{h}}V_{\rm eff}\simeq0$ and hence 
 the equation of motion (with $-V_{\rm eff}$) for $h-\bar{h}$ only admits a trivial solution $h-\bar{h}\simeq$(constant).
Under the above approximation, the phase transition is controlled by one-dimensional potential $V_{\rm eff}(h,h;\,\mu,T)$,
  which allows a qualitative discussion.
The one-dimensional potential reads
\begin{align}
V_{\rm eff}(h,h;\,\mu,T)&=\frac{1}{4}\lambda^2(h^2-v^2)^2
\label{tree2}\\
&+\frac{1}{64\pi^2}\sum_j M_{Bj}^4\left(\log\frac{M_{Bj}^2}{\mu^2}-\frac{3}{2}\right)-\frac{1}{64\pi^2}\sum_j M_{Fj}^4\left(\log\frac{M_{Fj}^2}{\mu^2}-\frac{3}{2}\right)
\label{one-loop2}\\
&+\frac{T^4}{2\pi^2}\sum_jJ_B(\frac{M_{Bj}^2}{T^2})-\frac{T^4}{2\pi^2}\sum_jJ_F(\frac{M_{Fj}^2}{T^2})
\label{finite-t2}
\end{align}
where $M_{Bj}^2$ are now obtained by diagonalizing
\begin{align}
{\cal M}^2_{\phi\bar{\phi}}&=\begin{pmatrix} % or pmatrix or bmatrix or Bmatrix or ...
        (4g_{B-L}^2+\frac{1}{2}\lambda^2)h^2  &  (-4g_{B-L}^2+\lambda^2)h^2-\frac{1}{2}\lambda^2 v^2 \\
          & (4g_{B-L}^2+\frac{1}{2}\lambda^2)h^2 \\
      \end{pmatrix}
      +  \frac{T^2}{16}\begin{pmatrix} % or pmatrix or bmatrix or Bmatrix or ...
      32g_{B-L}^2+2Y_{Mi}^2+4\lambda^2 & 0 \\
      0 &  32g_{B-L}^2+4\lambda^2 \\
   \end{pmatrix}
 \\
{\cal M}^2_{a\bar{a}}&=
      \frac{1}{2}\lambda^2\begin{pmatrix} % or pmatrix or bmatrix or Bmatrix or ...
        h^2  & v^2 \\
          & h^2 \\
      \end{pmatrix}
+   \frac{T^2}{16}\begin{pmatrix} % or pmatrix or bmatrix or Bmatrix or ...
      32g_{B-L}^2+2Y_{Mi}^2+4\lambda^2 & 0 \\
      0 &  32g_{B-L}^2+4\lambda^2 \\
   \end{pmatrix}
 \end{align}
 and also from ${\cal M}^2_S=\lambda^2h^2+\frac{1}{4}T^2\lambda^2$,
 ${\cal M}_{N_i^c}^2=\frac{1}{2}Y_{Mi}^2h^2+\frac{T^2}{8}(g_{B-L}^2+2Y_{Mi}^2)+\frac{3}{8}T^2g_{B-L}^2$ (2 duplications for each),
 $({\cal M}_X^2)^L=8g_{B-L}^2h^2+12g_{B-L}^2T^2$,
 and $({\cal M}_X^2)^T=8g_{B-L}^2h^2$ (2 duplications), while the MSSM particles become irrelevant.
One might guess that increasing $g_{B-L}$ and decreasing $\lambda$ 
 enhances the order of phase transition and hence the amount of latent heat,
 because the quartic coupling for $h$ is mostly $\lambda$, and the field-dependent mass for bosons 
 $\phi,\bar{\phi},X_\mu$ (which provides $h^3$ term in high-$T$ expansion)
 depends on $g_{B-L}^2h^2$ times a big factor 4 or 8.
However, increasing $g_{B-L}$ also enhances the thermal mass for these bosons (except for the transverse component of $X_\mu$), which diminishes their impact on the finite temperature effective potential.
Therefore, we expect that the amount of latent heat (which is related to $\alpha_\theta(T_{\rm n})$ in the next section)
 is maximized for $\lambda\to0$ and for some moderate value of $g_{B-L}$.
On the other hand, $Y_{Mi}$ is expected to have a weaker impact on the latent heat
 because it only appears in the field-dependent mass for $N_i^c$ and is not accompanied by a big factor.
All these expectations will be confirmed by the numerical study in the next section.
\\

\section{$U(1)_{B-L}$-breaking Phase Transition}

\subsection{$O(3)$-symmetric Euclidean Action}

We calculate the $O(3)$-symmetric Euclidean action~\cite{Linde:1980tt,Linde:1981zj} for a high-temperature $U(1)_{B-L}$-breaking phase transition 
from the metastable vacuum $(h,\bar{h})=(0,0)$ to an absolute vacuum where $(h,\bar{h})\neq(0,0)$.
Although we have seen in Section~2.3 that the potential is nearly symmetric with respect to $h$ and $\bar{h}$,
 we still consider a multi-field phase transition regarding $h$ and $\bar{h}$ as being independent.
To compute the $O(3)$-symmetric Euclidean action for a multi-field phase transition,
 we use CosmoTransitions~\cite{Wainwright:2011kj}.
From the action computed, we derive 
 the nucleation temperature, $T_{\rm n}$,
 the ratio of the trace anomaly divided by 4 over the radiation energy density of the symmetric phase at the nucleation temperature, $\alpha_\theta(T_{\rm n})$, and
 the speed of the phase transition at the nucleation temperature, $\beta_{\rm n}$.
They are defined as follows:

Let $S_E(T)$ denote the Euclidean action.
The tunneling rate per volume at temperature $T$ is $\Gamma(T)=A(T)e^{-S_E(T)/T}$, 
 where $A(T)$ is a factor with milder $T$-dependence than $e^{-S_E(T)/T}$.
The nucleation temperature $T_{\rm n}$ satisfies $H_{\rm n}^4 = A(T_{\rm n})e^{-S_E(T_{\rm n})/T_{\rm n}}$, where $H_{\rm n}$ denotes the Hubble rate at $T=T_{\rm n}$ in the symmetric phase.
We estimate $A(T_{\rm n})$ as $A(T_{\rm n})\sim T_{\rm n}^4$, and further approximate $T_{\rm n}$ by the $U(1)_{B-L}$-breaking VEV $v$. 
We estimate $H_{\rm n}$ as $H_{\rm n}^2\sim g_*\frac{\pi^2}{30}v^4\frac{1}{3M_*^2}$
  where $M_*$ is the reduced Planck mass and $g_*=255$ is the effective relativistic degrees of freedom of the SUSY $U(1)_{B-L}$ model including $\Phi,\overline{\Phi},S$ fields.
(We note that since we will find $\lambda/\sqrt{2} \ll T_{\rm n}/v$ in all our benchmarks,
 it is consistent to neglect the impact of soft SUSY breaking due to the $F$-term VEV of $S$ in the symmetric phase,
 on the calculation of the radiation energy density in that phase.)
Thus, $T_{\rm n}$ satisfies the relation
\bea
\frac{S_E(T_{\rm n})}{T_{\rm n}}\sim -\log\frac{(g_*\frac{\pi^2}{30})^2 v^4}{9M_*^4}.
\label{tnestimate}
\eea
For example, when $v=100$~TeV, the right-hand side of Eq.~(\ref{tnestimate}) equals 117, and when $v=1000$~TeV, it equals 107.
In the following analysis, we fix the right-hand side of Eq.~(\ref{tnestimate}) at 117.
We comment that we have computed the temperature at which the number of bubbles per Hubble horizon
 $N(T)=\int_{T_c}^T{\rm d}T'\frac{-1}{T'}\frac{\Gamma(T')}{H(T')^4}$ equals one, for the case with the largest supercooling among our benchmarks, by using Eqs.~(\ref{meta}),(\ref{stnum}), and we have found that this temperature agrees with $T_{\rm n}$ estimated by Eq.~(\ref{tnestimate}) with negligible discrepancy.
$\alpha_{\theta}(T_{\rm n})$ is given by
\bea
\alpha_{\theta}(T_{\rm n}) \ = \ \frac{1}{g_*\frac{\pi^2}{30}T^4}\left.\left(-\frac{T}{4}\frac{\partial \Delta V}{\partial T}+\Delta V\right)\right\vert_{T=T_{\rm n}}, \ \ \ \ \ 
\Delta V = V\vert_{\rm symmetric \ phase} - V\vert_{\rm broken \ phase}.
\label{alphaexp}
\eea
$\beta_{\rm n}$ satisfies
\bea
\beta_{\rm n}=-\left.\frac{{\rm d}}{{\rm d}t}\left(\frac{S_E(T)}{T}\right)\right\vert_{T=T_{\rm n}}=-\left.H_{\rm n}T\frac{{\rm d}}{{\rm d}T}\left(\frac{S_E(T)}{T}\right)\right\vert_{T=T_{\rm n}}.
\label{betaexp}
\eea

As with Section~2.3, we take the renormalization scale at $\mu=v$, which does not generate a large logarithm because $v$ is the only mass scale in the model.
We fix the right-hand side of Eq.~(\ref{tnestimate}) at 117, thereby neglecting its logarithmic dependence on $v/M_*$.
With the above choices,
 a quantity with mass dimension $n$ scales with $v^n$.
In particular, $T_{\rm n}$ scales with $v$, and so we present $T_{\rm n}/v$ in the plots.

\begin{figure}[H]
  \begin{center}
    \includegraphics[width=80mm]{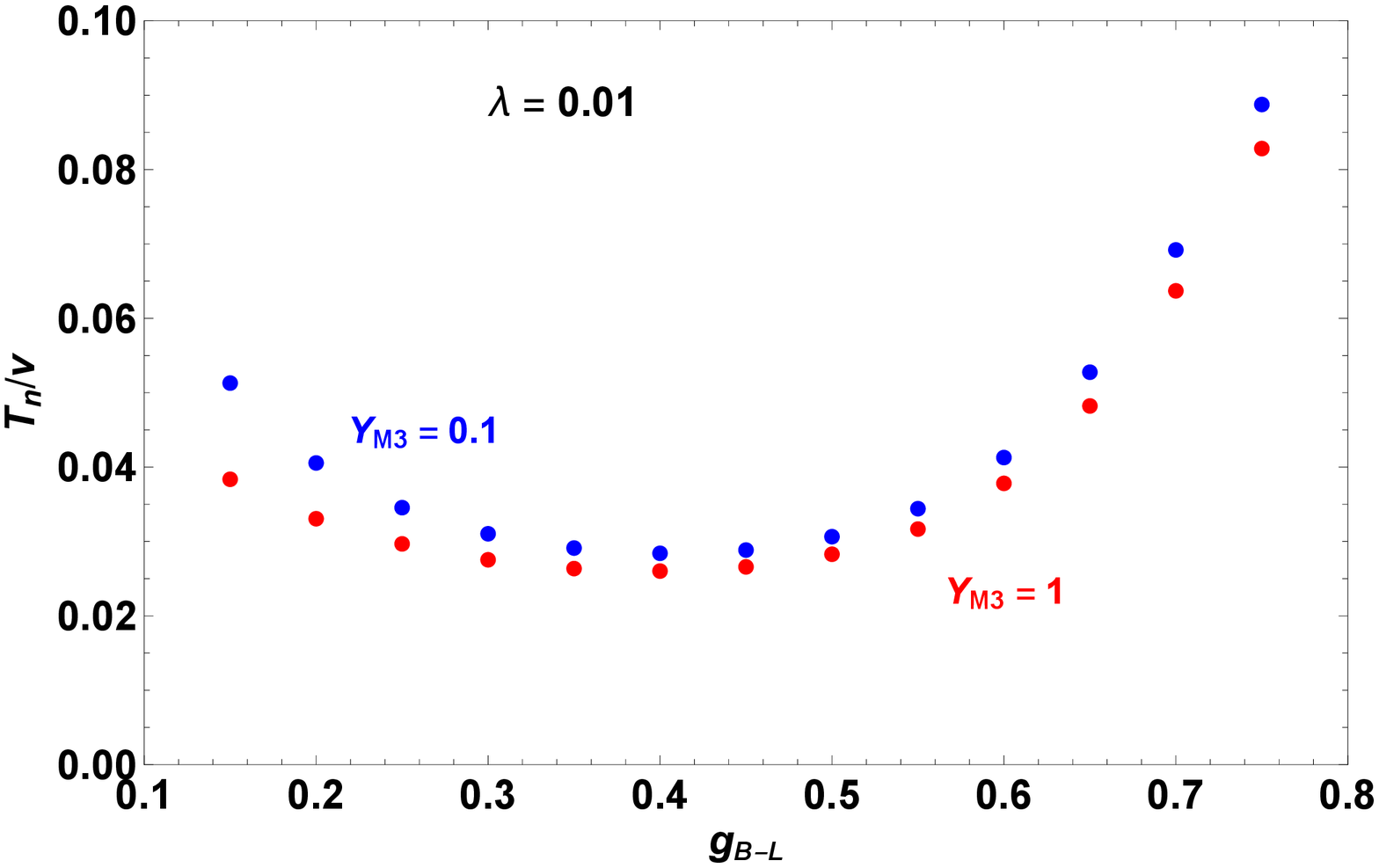}
    \includegraphics[width=80mm]{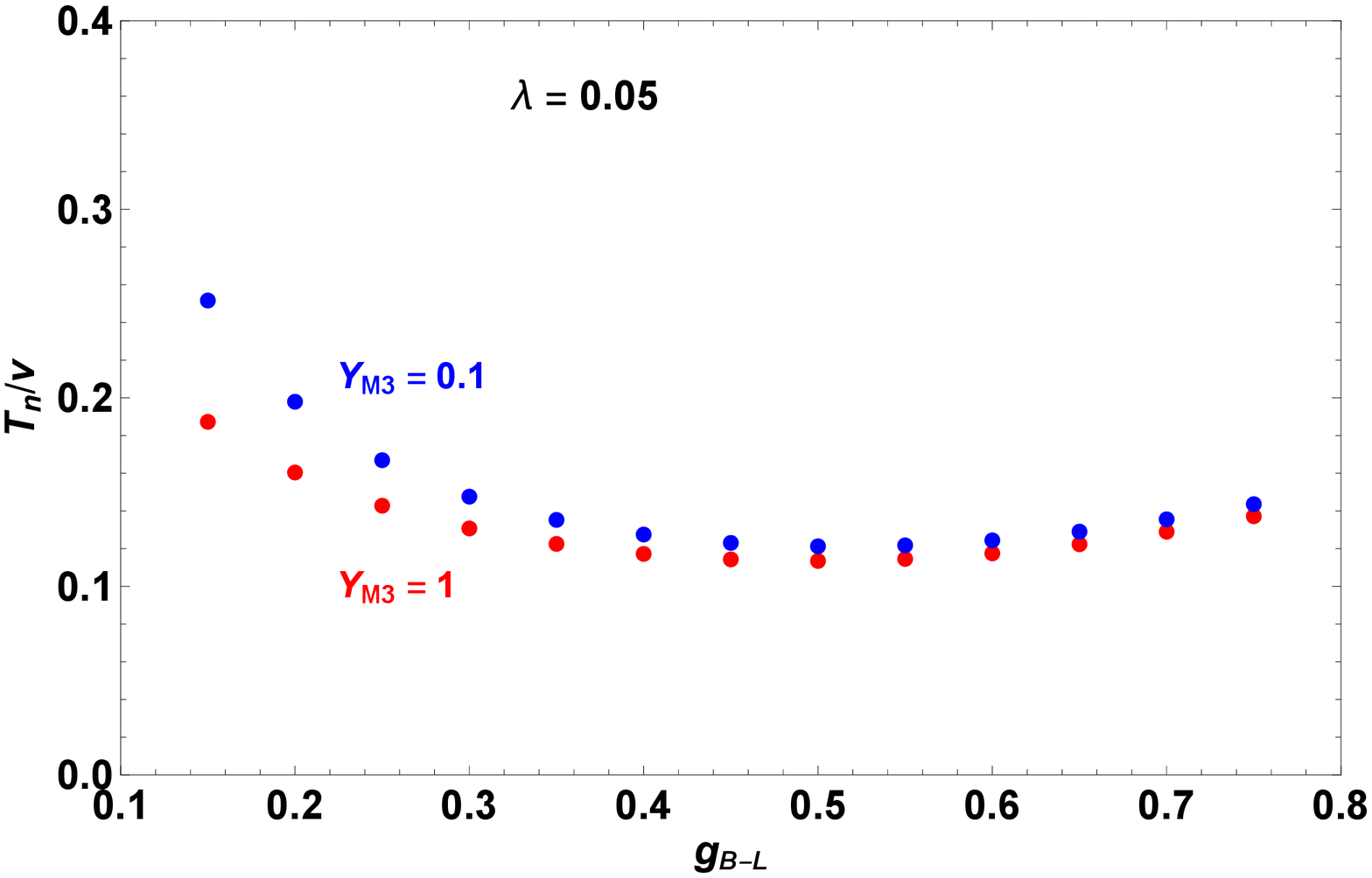}
    \caption{
    The nucleation temperature $T_{\rm n}$ evaluated from Eq.~(\ref{tnestimate}) by fixing the right-hand side at about 117,
    for various values of $U(1)_{B-L}$ gauge coupling $g_{B-L}$ and for $\lambda=0.01,\,0.05$ and $Y_{M3}=1,\,0.1$. 
    (We fix $Y_{M1}=Y_{M2}=0$.)
    }
    \label{TnPlot}
  \end{center}
\end{figure}

In Fig.~\ref{TnPlot}, we plot $g_{B-L}$-dependence of the nucleation temperature $T_{\rm n}$,
 for $\lambda=0.01,\,0.05$ and $Y_{M3}=1,\,0.1$, with $Y_{M1}=Y_{M2}=0$.
We find that $T_{\rm n}/v$ has little dependence on $Y_{M3}$, and is much affected by $\lambda$.

\begin{figure}[H]
  \begin{center}
    \includegraphics[width=80mm]{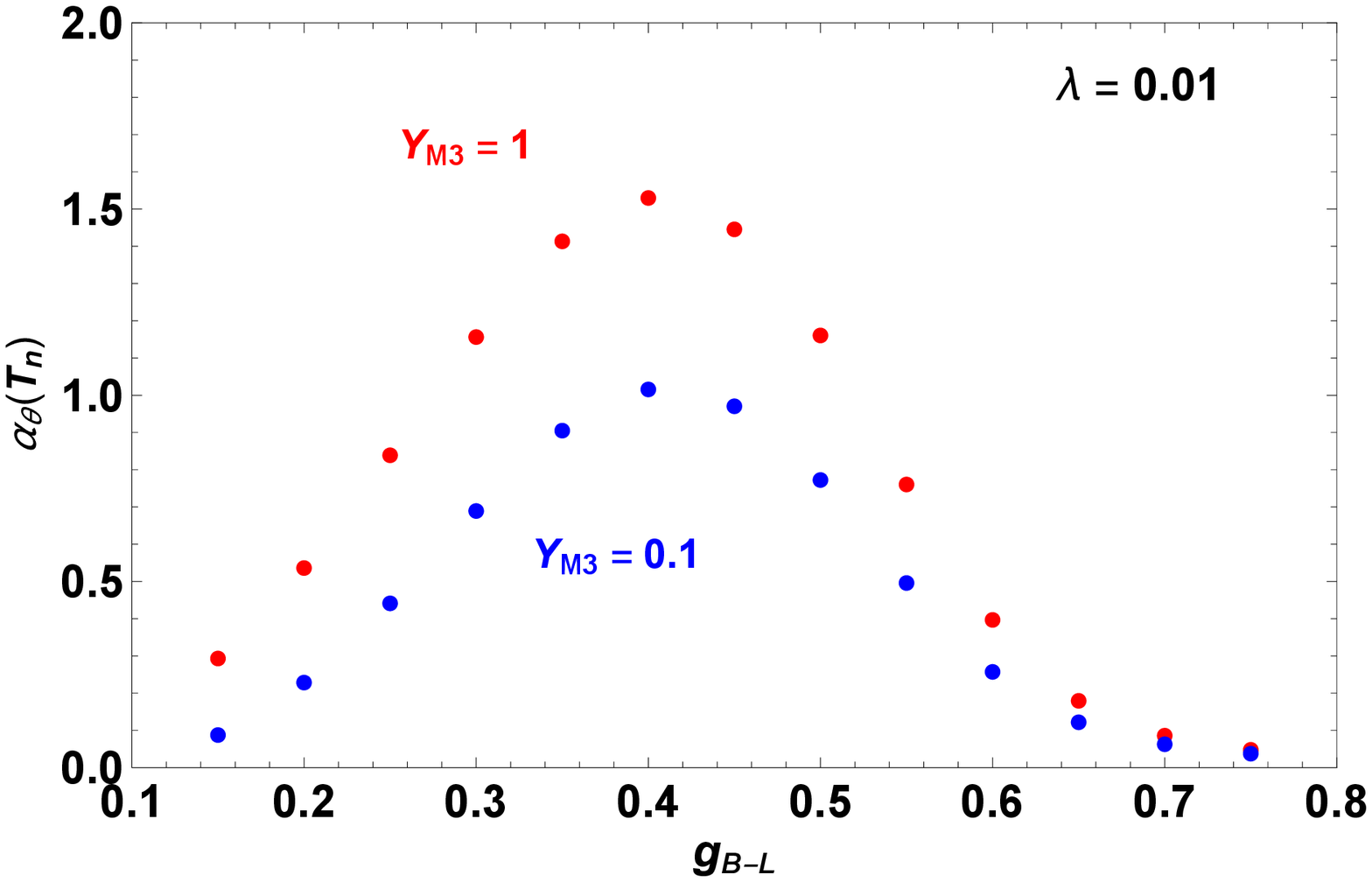}
    \includegraphics[width=80mm]{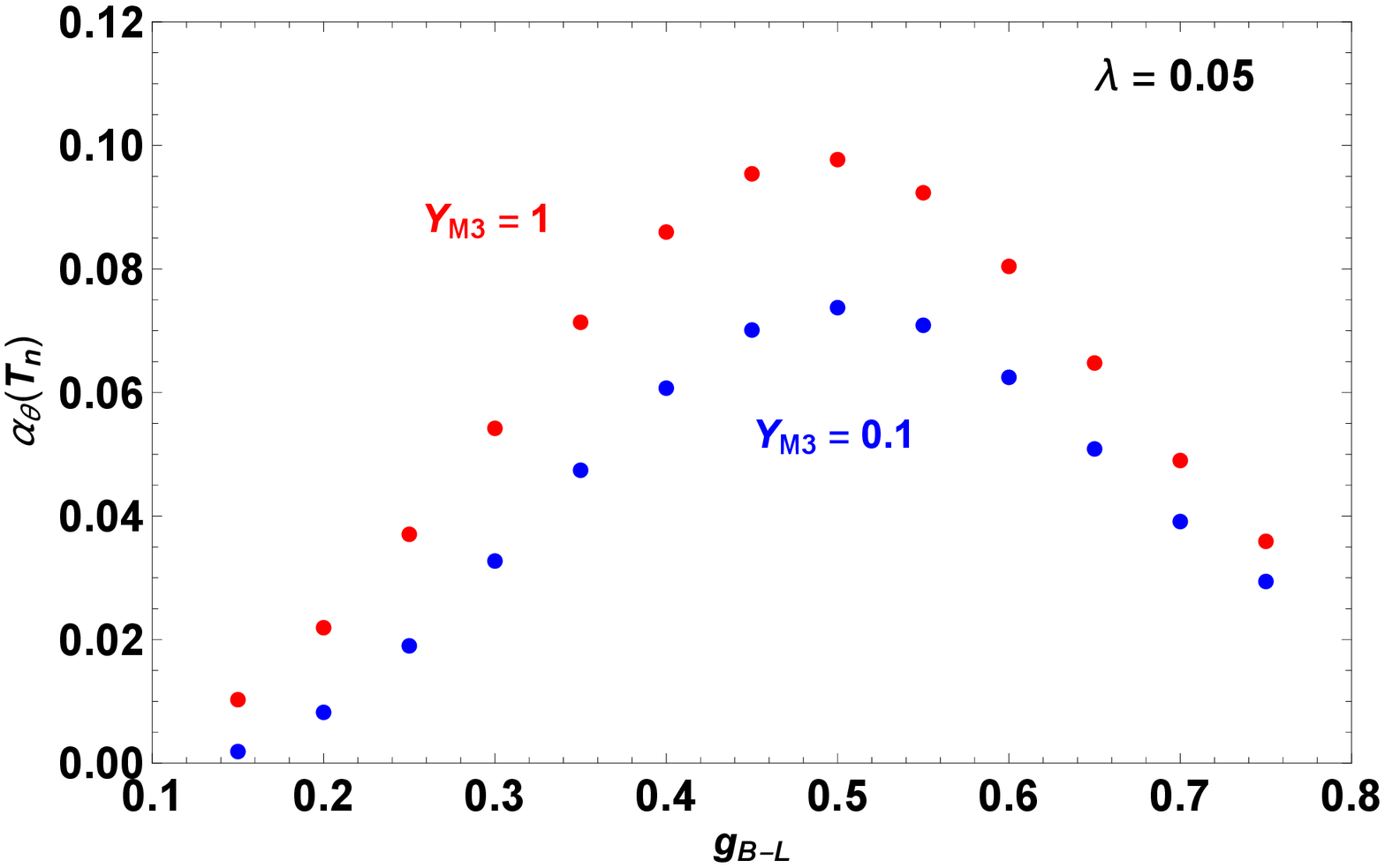}
    \caption{
   The trace anomaly divided by 4 over the radiation energy density $\alpha_\theta(T_{\rm n})$ Eq.~(\ref{alphaexp}),
    for various values of $U(1)_{B-L}$ gauge coupling $g_{B-L}$ and for $\lambda=0.01,\,0.05$ and $Y_{M3}=1,\,0.1$. 
    (We fix $Y_{M1}=Y_{M2}=0$.)
    }
    \label{alphaPlot}
  \end{center}
\end{figure}

In Fig.~\ref{alphaPlot}, we plot $g_{B-L}$-dependence of the trace anomaly divided by 4 over the radiation energy density 
 $\alpha_\theta(T_{\rm n})$,
 for $\lambda=0.01,\,0.05$ and $Y_{M3}=1,\,0.1$, with $Y_{M1}=Y_{M2}=0$.
$\alpha_\theta(T_{\rm n})$ is significantly enhanced for $\lambda=0.01$ compared to the case with $\lambda=0.05$.
Interestingly, $\alpha_\theta(T_{\rm n})$ is maximized at $g_{B-L}\simeq0.4$ when $\lambda=0.01$, 
and at $g_{B-L}\simeq0.5$ when $\lambda=0.05$.
The dependence on $Y_{M3}$ is quite mild compared to those on $g_{B-L}$ and $\lambda$.

\begin{figure}[H]
  \begin{center}
    \includegraphics[width=80mm]{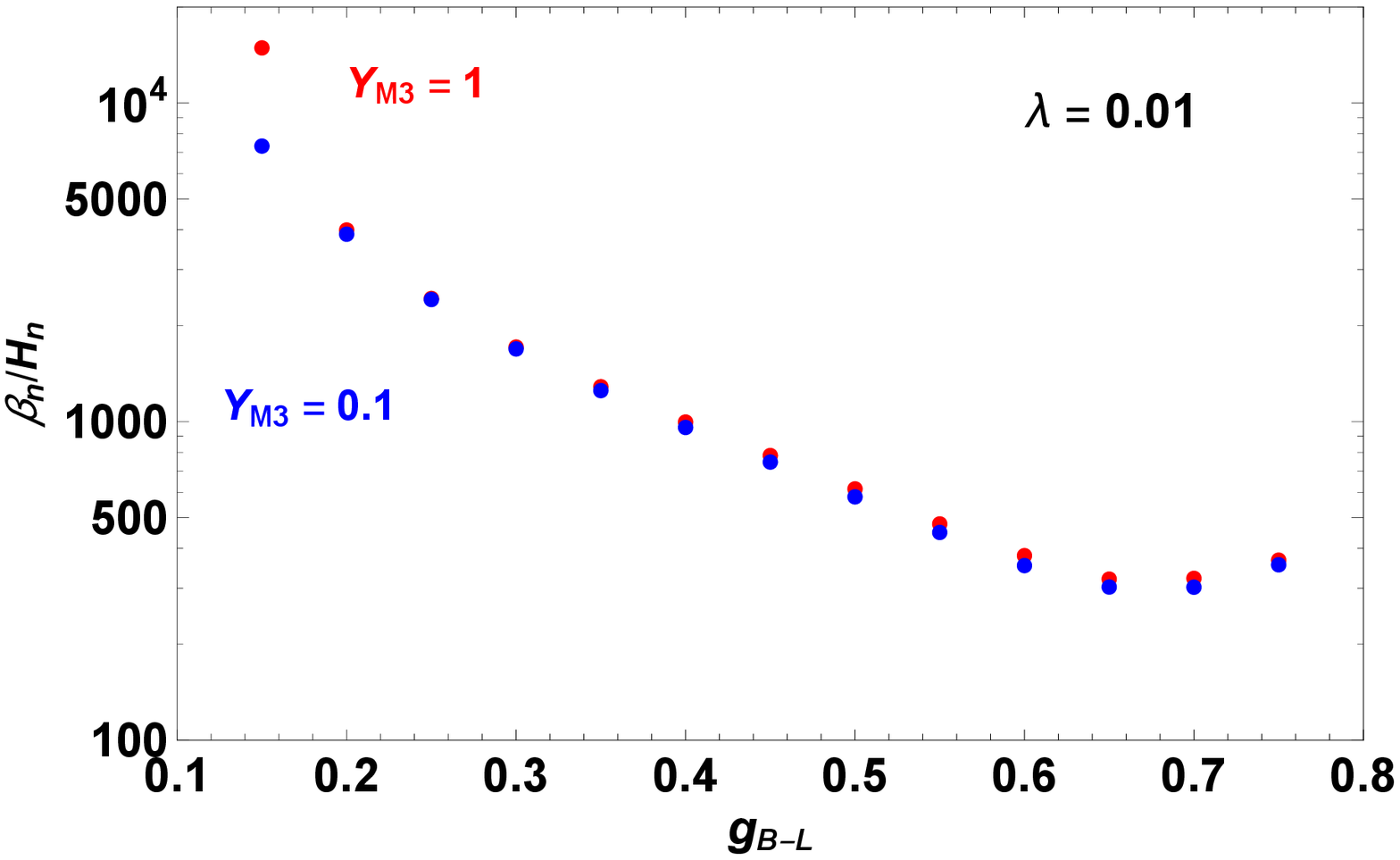}
    \includegraphics[width=80mm]{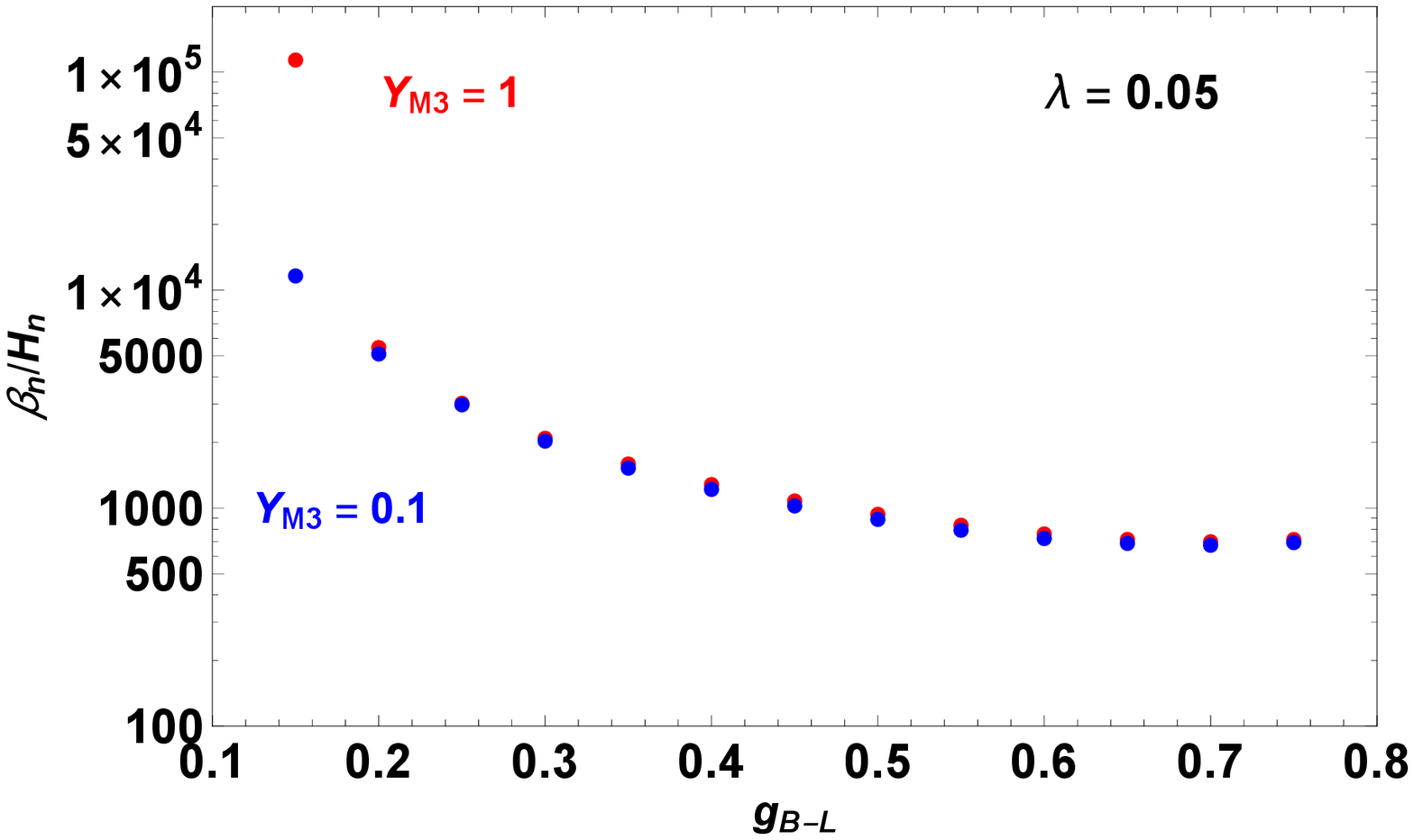}
    \caption{
   The speed of phase transition in units of the Hubble rate at the nucleation temperature
   $\beta_{\rm n}/H_{\rm n}$ Eq.~(\ref{betaexp}),
    for various values of $U(1)_{B-L}$ gauge coupling $g_{B-L}$ and for $\lambda=0.01,\,0.05$ and $Y_{M3}=1,\,0.1$. 
    (We fix $Y_{M1}=Y_{M2}=0$.)
    }
    \label{betaPlot}
  \end{center}
\end{figure}

In Fig.~\ref{betaPlot}, we plot $g_{B-L}$-dependence of the speed of phase transition in units of the Hubble rate at the nucleation temperature
 $\beta_{\rm n}/H_{\rm n}$ in logarithm,
 for $\lambda=0.01,\,0.05$ and $Y_{M3}=1,\,0.1$, with $Y_{M1}=Y_{M2}=0$.
$\beta_{\rm n}/H_{\rm n}$ is exponentially enhanced for small values of $g_{B-L}$.
In contrast, the dependence on $Y_{M3}$ is negligible, except for $g_{B-L}=0.15$.

\begin{figure}[H]
  \begin{center}
    \includegraphics[width=80mm]{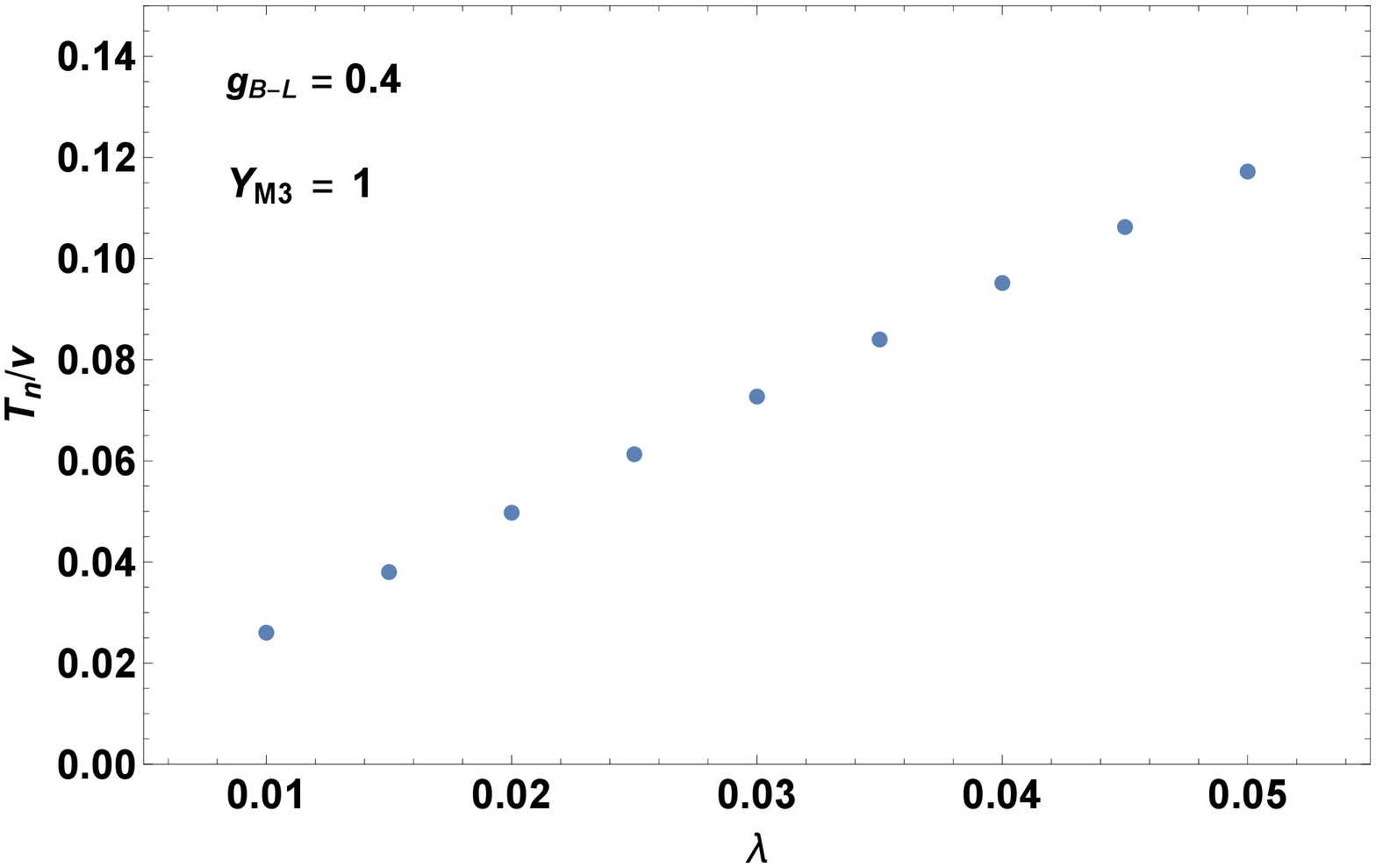}
    \includegraphics[width=80mm]{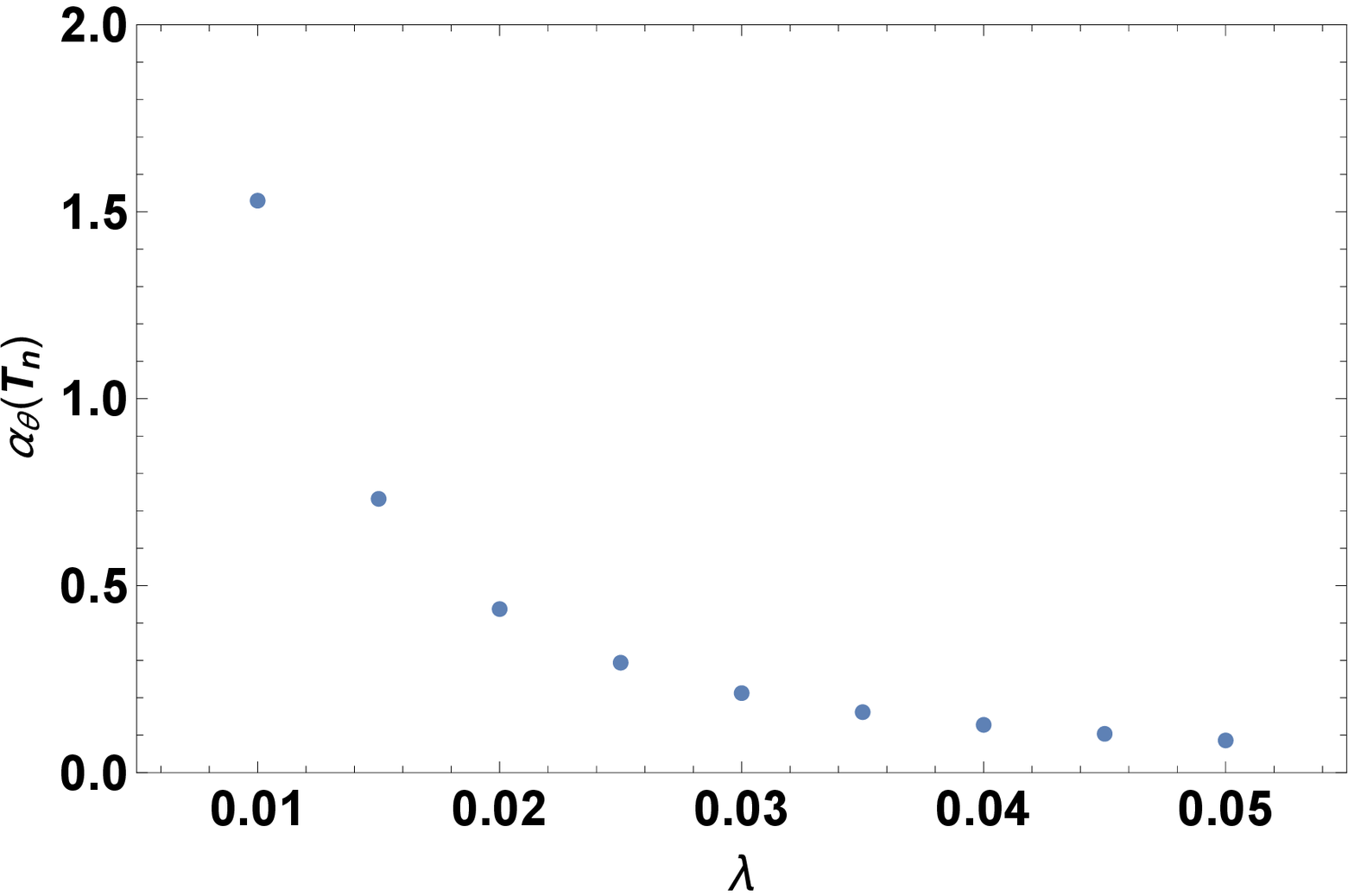}    
    \includegraphics[width=80mm]{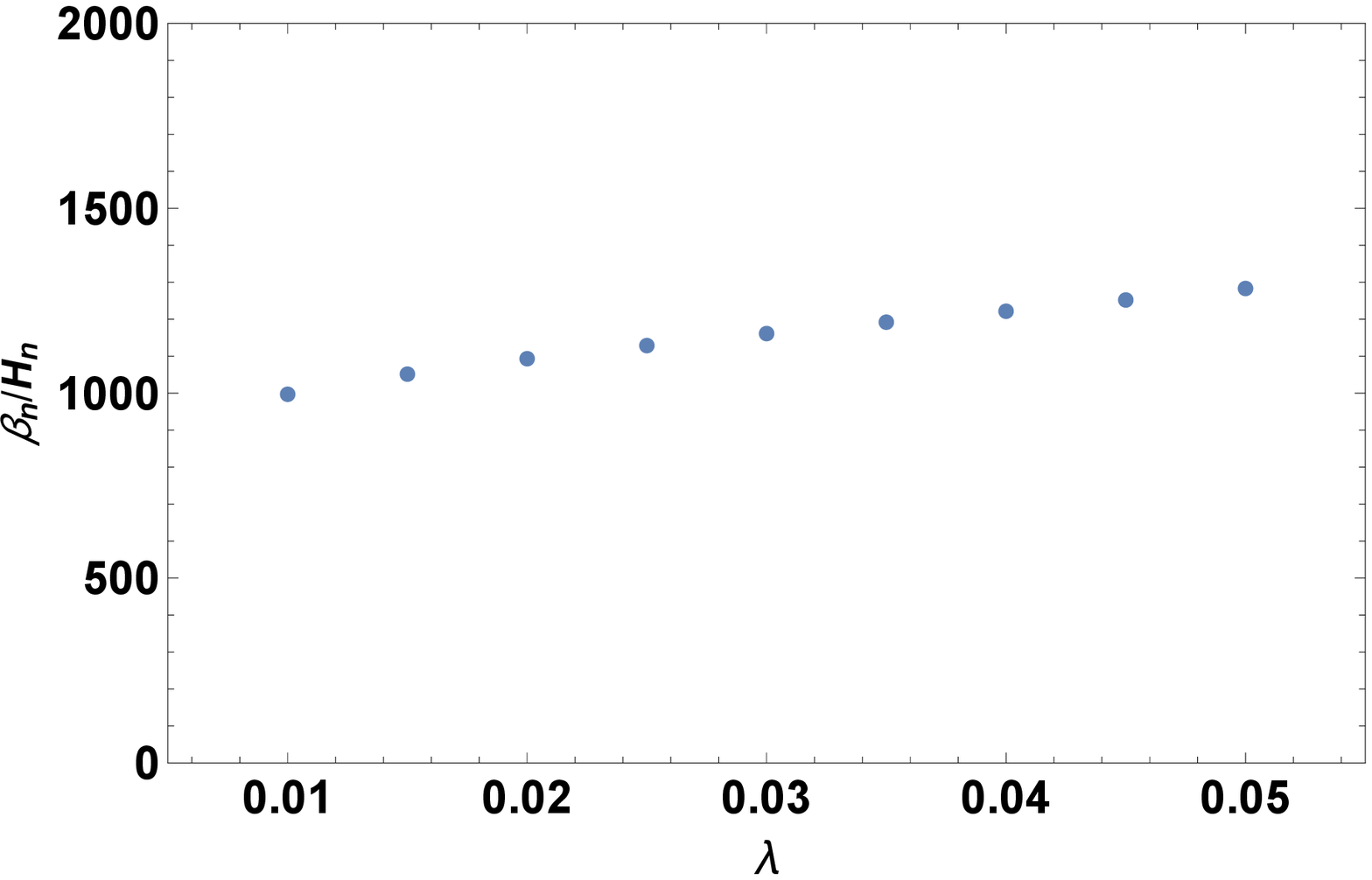}
    \caption{
    The dependence of $T_{\rm n},\,\alpha_\theta(T_{\rm n}),\,\beta_{\rm n}/H_{\rm n}$ on $\lambda$ for $g_{B-L}=0.4$ and $Y_{M3}=1$.
    }
    \label{lambdadep}
  \end{center}
\end{figure}

Finally, we study how the above quantities vary with $\lambda$.
We concentrate on an interesting case where $g_{B-L}=0.4$ and $Y_{M3}=1$, which has given the largest 
$\alpha_\theta(T_{\rm n})$ in the above plots when $\lambda=0.01$.
The dependence of $T_{\rm n},\,\alpha_\theta(T_{\rm n}),\,\beta_{\rm n}/H_{\rm n}$ on $\lambda$ for $g_{B-L}=0.4$ and $Y_{M3}=1$
is found in Fig.~\ref{lambdadep}.
It is observed that $T_{\rm n}$ increases linearly with $\lambda$, 
 and $\beta_{\rm n}/H_{\rm n}$ has almost no dependence on $\lambda$,
 while $\alpha_\theta(T_{\rm n})$ decreases much rapidly with $\lambda$.
\label{benchmarks}
\\

%%%%%%%%%%%%%%%%%%%%
\subsection{Percolation}

When $\alpha_\theta(T)>1$, vacuum energy stored in the meta-stable vacuum 
 causes inflation of meta-stable-vacuum region,
 which hinders the percolation of absolute-vacuum bubbles~\cite{Ellis:2018mja} 
 (first considered for zero-temperature, quantum phase transitions in \cite{Guth:1982pn}).
In this subsection, we focus on the benchmark of $(\lambda,~g_{B-L},~Y_{M3})=(0.01,~0.4,~1)$,
 which has given $\alpha_\theta(T_{\rm n})=1.5$,
 the largest $\alpha_\theta(T_{\rm n})$ among the benchmarks of Section~\ref{benchmarks},
 and show that the percolation is completed despite large vacuum energy of the meta-stable vacuum.
We can then infer that the percolation is also completed in benchmarks with smaller $\alpha_\theta(T_{\rm n})$.

We define the energy such that the vacuum energy of the absolute (broken) vacuum is zero, 
 to be in agreement with the observed almost-zero cosmological constant.
Let $T$ denote the temperature of the radiation in the meta-stable (symmetric) vacuum.
The energy density of the meta-stable vacuum is
\footnote{
Again, we neglect the impact of soft SUSY breaking due to the $F$-term VEV of $S$, since $\lambda/\sqrt{2}\ll T_{\rm n}/v$.
}
\bea
\rho_{\rm meta}(T)&=&g_*\frac{\pi^2}{30}T^4+\rho_0, \ \ \ \ \ \ \ \ \ \ \ \ \ \ \ \rho_0 \ = \ \frac{1}{4}\lambda^2\,v^4,
\label{meta}
\eea
 where $g_*=255$ is the effective relativistic degrees of freedom of the minimal SUSY $U(1)_{B-L}$ model,
 and $\rho_0=\frac{1}{4}\lambda^2\,v^4$ comes the $F$-term VEV of $S$ field.

We have found numerically that 
 the temperature-dependence of the $O(3)$-symmetric Euclidean action in the benchmark of
 $(\lambda,~g_{B-L},~Y_{M3})=(0.01,~0.4,~1)$ is well approximated by
\bea
\frac{S_E(T)}{T}&=&2.43\times10^6 \left(\frac{T}{v}-0.0182\right)^2 -32.4 \ \ \ \ \ {\rm for} \ T/v>0.0219.
\label{stnum}
\eea
Here $T/v=0.0219$ is the temperature at which the potential barrier disappears (then $S_E\simeq0$).
Note that $S_E(T)/T$ is monotonic in this range.

\begin{figure}[H]
  \begin{center}
    \includegraphics[width=80mm]{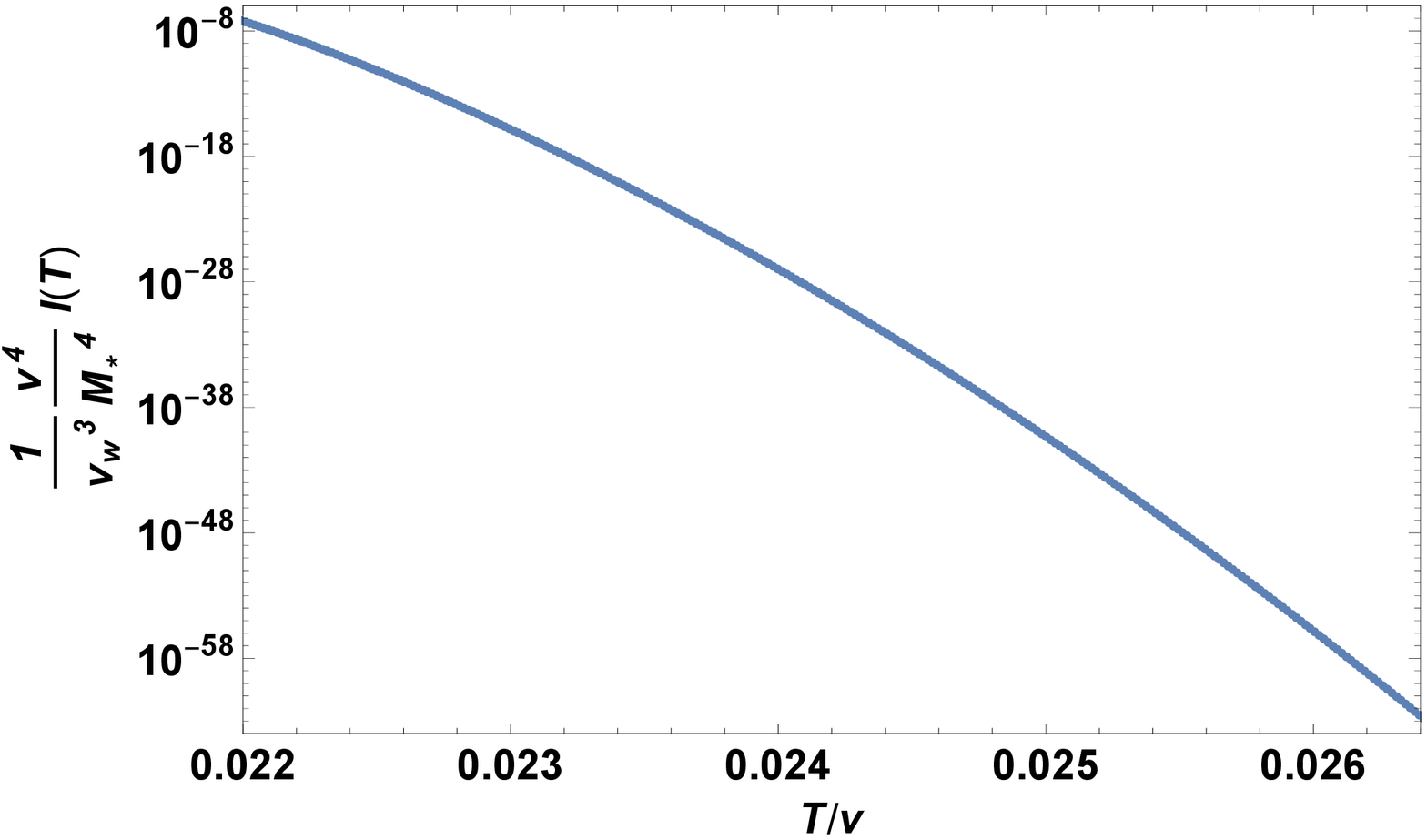}
    \includegraphics[width=80mm]{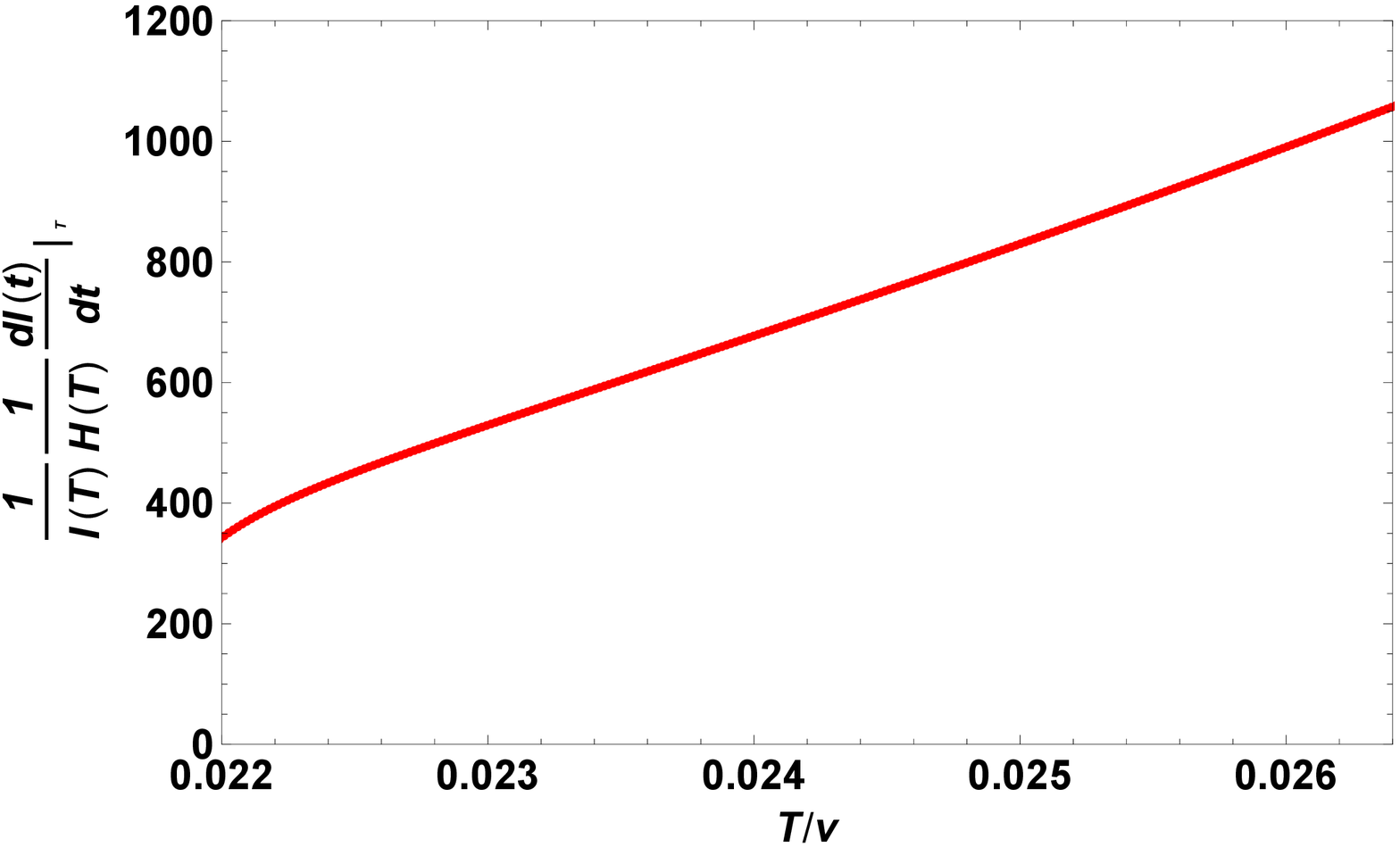}
    \caption{
    Fraction of absolute-vacuum bubbles when their overlaps are neglected, $I(T)$,
     and the ratio of its time derivative over the Hubble rate over itself, $\frac{1}{I(T)}\frac{1}{H(T)}\frac{{\rm d}I}{{\rm d}t}|_T$,
     in the benchmark of $(\lambda,~g_{B-L},~Y_{M3})=(0.01,~0.4,~1)$.
             }
    \label{percolation}
  \end{center}
\end{figure}

Now we study the probability of finding a point in the meta-stable vacuum, $P(t)$.
It is given by~\cite{Guth:1979bh,Guth:1981uk}
\footnote{
Unless $v_{\rm w}=1$, the bubble expansion breaks the homogeneity of meta-stable-vacuum region
 and this region is not described by Friedmann-Robertson-Walker (FRW) metric.
Nevertheless, we assume that FRW metric gives a good approximation even for $v_{\rm w}<1$.
}
\bea
P(t) = e^{-I(t)}, \ \ \ \ \ \ \ I(t) = \frac{4\pi}{3}\int_{t_c}^t{\rm d}t' \, \Gamma(t')a(t')^3\left(\int_{t'}^t{\rm d}\tilde{t}\,\frac{v_{\rm w}}{a(\tilde{t})}\right)^3,
\eea
where $I(t)$ is the fraction of absolute-vacuum bubbles when their overlaps are neglected.
$t_c$ is the time corresponding to the critical temperature,
  $\Gamma(t)$ denotes the tunneling rate per volume, and $v_{\rm w}$ denotes the speed of the bubble wall.
For later use, we also give the time derivative of $I(t)$,
\bea
\frac{{\rm d}I(t)}{{\rm d}t} \ = \ 4\pi\frac{v_{\rm w}}{a(t)}\int_{t_c}^t{\rm d}t' \, \Gamma(t')a(t')^3
\left(\int_{t'}^t{\rm d}\tilde{t}\,\frac{v_{\rm w}}{a(\tilde{t})}\right)^2.
\eea
Since there is no entropy production in the meta-stable vacuum, we can rewrite $I(t)$ and its time derivative in terms of $T$ as
\bea
I(T)&=&\frac{4\pi}{3}\int_{T_c}^T{\rm d}T'\,
\frac{-1}{H(T')}\frac{1}{T'^4}\Gamma(T')\left(\int_{T'}^T{\rm d}\tilde{T}\,\frac{-v_{\rm w}}{H(\tilde{T})}\right)^3,
\label{it}\\
\left.\frac{{\rm d}I(t)}{{\rm d}t}\right|_T&=&
4\pi Tv_{\rm w}\int_{T_c}^T{\rm d}T'\,
\frac{-1}{H(T')}\frac{1}{T'^4}\Gamma(T')\left(\int_{T'}^T{\rm d}\tilde{T}\,\frac{-v_{\rm w}}{H(\tilde{T})}\right)^2
\eea
 where $H(T)$ is the Hubble rate of the meta-stable vacuum given from Eq.~(\ref{meta}).
A criterion for the completion of the percolation is that~\cite{Ellis:2018mja,Turner:1992tz}
 there is a temperature $T_p$ at which $I(T_p)=0.34$ and the physical volume of meta-stable-vacuum region decreases with time, i.e.,
 \bea
 0&>&\frac{1}{a(t)^3 P(t)}\left.\frac{{\rm d}}{{\rm d}t}\{ a(t)^3 P(t)\}\right|_{T_p} \ =\  
 H(T_p)\left\{3-\frac{1}{H(T_p)}\left.\frac{{\rm d}I(t)}{{\rm d}t}\right|_{T_p}\right\}.
 \eea
To see if the above criterion is fulfilled in the benchmark of $(\lambda,~g_{B-L},~Y_{M3})=(0.01,~0.4,~1)$,
 we compute $I(T)$ and $\frac{1}{H(T)}\frac{{\rm d}I}{{\rm d}t}|_T$ from Eqs.~(\ref{meta}),(\ref{stnum}) with
 the formula $\Gamma(T)=T^4\{S_E(T)/2\pi T\}^{3/2}$ $e^{-S_E(T)/T}$, and plot $I(T)$
 in units of $v_{\rm w}^3(M_*^4/v^4)$ and $\frac{1}{I(T)}\frac{1}{H(T)}\frac{{\rm d}I}{{\rm d}t}|_T$
 in Fig.~\ref{percolation}.
From the left panel of Fig.~\ref{percolation}, we see that if 10~TeV$\lesssim v\lesssim1000$~TeV 
 (which is phenomenologically relevant because it safely satisfies the collider constraint on the $U(1)_{B-L}$ gauge boson
  and it may lead to a gravitational wave spectrum whose peak is covered by Cosmic Explorer and Einstein Telescope) 
  and thus $10^{57}\gtrsim M_*^4/v^4\gtrsim 10^{49}$,
 the relation $I(T_p)=0.34$ is fulfilled somewhere in the range $0.025<T_p/v<0.026$, for $v_{\rm w}>O(0.1)$.
Furthermore, the right panel manifests that the relation $\frac{1}{H(T)}\frac{{\rm d}I}{{\rm d}t}|_{T}>200\cdot I(T)$ holds for 
 $0.022<T/v<0.0264$, and hence $3-\frac{1}{H(T_p)}\frac{{\rm d}I}{{\rm d}t}|_{T_p}<0$ holds.
Therefore, in the interesting parameter range where 10~TeV$\lesssim v\lesssim1000$~TeV,
 the criterion for the completion of the percolation is fulfilled,
 namely, the phase transition ends by the coalescence of absolute-vacuum bubbles and the standard cosmology is recovered after that.

We crosscheck the above result using approximate expressions.
For a given temperature $T$, let $\Delta T$ denote a temperature difference
 for which $\frac{S_E(T+\Delta T)}{T+\Delta T}=\frac{S_E(T)}{T}+O(1)$.
Because $\Gamma(T)$ is a monotonic function with rapid $T$-dependence through the factor $e^{-S_E(T)/T}$,
 the integrals of Eqs.~(\ref{meta}),(\ref{stnum}) are dominated by the region around $T$
 and thus we can make the following approximations:
\bea
I(T)&\simeq&\frac{4\pi}{3}v_{\rm w}^3 \frac{\Delta T^4}{H(T)^4}\left(\frac{S_E(T)}{2\pi T}\right)^{3/2}e^{-S_E(T)/T},
\label{itapproximation}\\
\left.\frac{1}{H(T)}\frac{{\rm d}I(t)}{{\rm d}t}\right|_T
&\simeq&4\pi v_{\rm w}^3 \frac{\Delta T^3 T}{H(T)^4}\left(\frac{S_E(T)}{2\pi T}\right)^{3/2}e^{-S_E(T)/T}.
\label{ditapproximation}
\eea
We find numerically that taking
\bea
\Delta T&=&\frac{1.4}{\frac{{\rm d}}{{\rm d}T}\left(\frac{S_E(T)}{T}\right)}
\label{deltat}
\eea
  gives a good order-of-magnitude estimate in the range $0.0230<T/v<0.0264$.
By inserting Eqs.~(\ref{meta}),(\ref{stnum}),(\ref{deltat}) into Eq.~(\ref{itapproximation}), we see that
 the relation $I(T_p)=0.34$ is satisfied somewhere in the range $0.025<T_p/v < 0.026$
 for 10~TeV$\lesssim v\lesssim1000$~TeV and $v_{\rm w}>O(0.1)$.
By taking the ratio of Eqs.~(\ref{itapproximation}),(\ref{ditapproximation}) and inserting Eqs.~(\ref{meta}),(\ref{stnum}),(\ref{deltat}),
 we get
\bea
\left.\frac{1}{I(T)}\frac{1}{H(T)}\frac{{\rm d}I}{{\rm d}t}\right|_{T}&=&3\frac{T}{\Delta T} \ = \
 1.0\times10^7\cdot\frac{T}{v}\left(\frac{T}{v}-0.0182\right),
\eea
 which is much greater than $3/0.34$ in the range $0.0230<T/v<0.0264$,
 and hence the relation $3-\frac{1}{H(T_p)}\frac{{\rm d}I}{{\rm d}t}|_{T_p}<0$ is easily satisfied.
\\

%%%%%%%%%%%%%%
\subsection{Gravitational Waves}

We estimate gravitational waves generated from a $U(1)_{B-L}$-breaking phase transition in the early Universe.
In this subsection, we exclusively study the case with $\lambda=0.05$, 
 which gives $\alpha_\theta(T_{\rm n})<0.1$ (see the right panel of Fig.~\ref{alphaPlot}).
This selection is because the study on gravitational wave production in a strong phase transition $\alpha_\theta(T_{\rm n})>0.1$
 is currently under development (see, e.g., Refs.~\cite{Cutting:2019zws,Pol:2019yex}),
 while that in a weaker phase transition is relatively well established.

The sources of gravitational waves from a finite-temperature phase transition are 
 (i) the energy momentum tensor of scalar field in colliding bubbles,
 (ii) that of sound waves of a surrounding plasma,
 and (iii) that of magnetohydrodynamic turbulence of a surrounding plasma.
On the basis of the claim of Ref.~\cite{Bodeker:2017cim} that the bubble wall in plasma always reaches a constant velocity when the next-to-leading order friction is taken into account, and a claim of Ref.~\cite{Ellis:2019oqb} that the fraction of energy stored in the bubble wall over the vacuum energy released quickly decreases after the wall reaches the constant velocity,
we justifiably neglect source (i).
Hence, we only consider sources (ii) and (iii).

For source (ii), it is claimed in Ref.~\cite{Hindmarsh:2015qta} that the energy spectrum of gravitational waves generated by sound waves in a hot plasma
 in a phase transition with $\alpha_\theta(T_{\rm n})\lesssim0.1$ can be expressed as
 (we rewrite the formula for gravitational wave energy over the critical density we observe today)
\bea
\frac{{\rm d}\Omega_{\rm sound}(k)h^2}{{\rm d}\log k} \ = \ 3H_{\rm n}L_{\rm f,n}\frac{1}{2\pi^2}(kL_{\rm f})^3\,(1+\overline{p}/\overline{\epsilon})^2\,\overline{U}_{\rm f}^4 \ \tilde{P}_{\rm gw}(kL_{\rm f})
\times 1.2\cdot10^{-5}\left(\frac{255}{g_*}\right)^{1/3},
\label{sw}
\eea
 where $L_{\rm f,n}$ is a typical length scale of fluid motion at the nucleation temperature, $L_{\rm f}$ is the redshifted value of $L_{\rm f,n}$ today,
 and $\tilde{P}_{\rm gw}$ is a function only of the product $kL_{\rm f}$.
$\overline{U}_{\rm f}$ is the enthalpy-weighted root mean square four-velocity of fluid at the nucleation temperature,
 and $1+\overline{p}/\overline{\epsilon}$ is the ratio of enthalpy over energy.
In this paper, we adopt Eq.~(\ref{sw}).
We further identify $L_{\rm f,n}$ with the mean bubble separation $(8\pi)^{1/3} v_{\rm w}/\beta_{\rm n}$~\cite{Enqvist:1991xw}
 ($v_{\rm w}$ denotes the bubble wall speed),
 and for $\tilde{P}_{\rm gw}$, we use a fitting of the simulation results in Ref.~\cite{Hindmarsh:2017gnf}, which has improved on earlier works~\cite{Hindmarsh:2013xza,Hindmarsh:2015qta}.
For $(1+\overline{p}/\overline{\epsilon})\overline{U}_{\rm f}^2$, we use a fitting formula for the ratio of bulk kinetic energy over vacuum energy $\kappa(\alpha_{\theta},\,v_{\rm w})$ derived in Ref.~\cite{Espinosa:2010hh}, and evaluate it as
\bea
(1+\overline{p}/\overline{\epsilon})\,\overline{U}_{\rm f}^2 \ = \ \frac{\alpha_\theta(T_{\rm n})}{1+\alpha_\theta(T_{\rm n})}\kappa\left(\alpha_\theta(T_{\rm n}),\,v_{\rm w}\right).
\eea
The calculation of the bubble wall speed $v_{\rm w}$ is beyond the scope of the current paper,
 and we simply assume various values of $v_{\rm w}$ that appear in the simulations of Ref.~\cite{Hindmarsh:2017gnf} and 
 evaluate gravitational wave spectrum in each case.

For source (iii), we estimate its contribution by the following formula in Ref.~\cite{Caprini:2015zlo}, 
 which is based on Refs.~\cite{Caprini:2009yp,Binetruy:2012ze}:
 \begin{align}
\frac{{\rm d}\Omega_{\rm turb}(k)h^2}{{\rm d}\log k} &= 3.35\times10^{-4}\frac{H_{\rm n}}{\beta_{\rm n}}
\left(\frac{\kappa_{\rm turb}\alpha_\theta(T_{\rm n})}{1+\alpha_\theta(T_{\rm n})}\right)^{\frac{3}{2}}\left(\frac{100}{g_*}\right)^{\frac{1}{3}}v_{\rm w}
\frac{(k/k_{\rm turb})^3}{(1+k/k_{\rm turb})^{\frac{11}{3}}\{1+4(k/H_{\rm n})(a_0/a_{\rm n})\}}
\label{turb}
\\
k_{\rm turb} \ &= \ 2\pi\times 2.7\times10^{-5}~{\rm Hz}\times \frac{1}{v_{\rm w}}\frac{\beta_{\rm n}}{H_{\rm n}}
\frac{T_{\rm n}}{100~{\rm GeV}}\left(\frac{g_*}{100}\right)^{\frac{1}{6}}
\end{align}
 where $a_0/a_{\rm n}$ is the red shift factor.
We estimate $\kappa_{\rm turb}$ aggressively as $\kappa_{\rm turb}=0.1\cdot\kappa(\alpha_{\theta},\,v_{\rm w})$
 following Ref.~\cite{Caprini:2015zlo}.

 The total gravitational wave spectrum is given by
 \bea
\frac{{\rm d}\Omega_{\rm gw}(k)h^2}{{\rm d}\log k} \ = \ 
\frac{{\rm d}\Omega_{\rm sound}(k)h^2}{{\rm d}\log k}+\frac{{\rm d}\Omega_{\rm turb}(k)h^2}{{\rm d}\log k}.
 \eea

 We comment that the relation on which the simulation of Ref.~\cite{Hindmarsh:2017gnf} relies, $H_{\rm n}L_{\rm f,n}> \overline{U}_{\rm f}$, is not satisfied in our benchmark.
This means that sound waves turn into turbulence in less than a Hubble time,
 which suppresses the sound waves' contribution to gravitational waves compared to Eq.~(\ref{sw}),
 and may enhance the turbulence's contribution compared to Eq.~(\ref{turb})~\cite{Ellis:2019oqb}.
Nevertheless, we use Eqs.~(\ref{sw}),(\ref{turb}) in the current analysis.

\begin{figure}[H]
  \begin{center}
    \includegraphics[width=80mm]{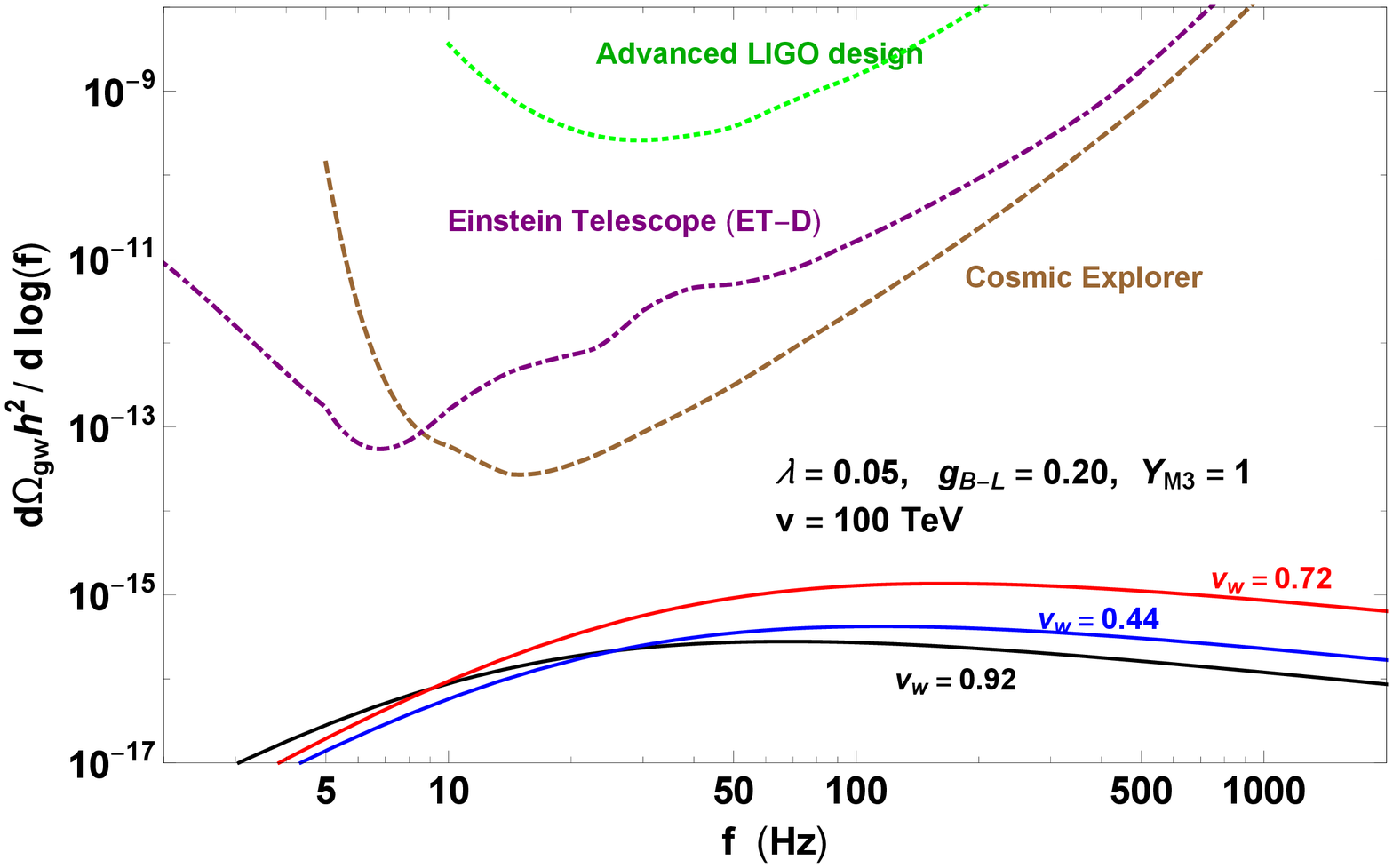}
    \includegraphics[width=80mm]{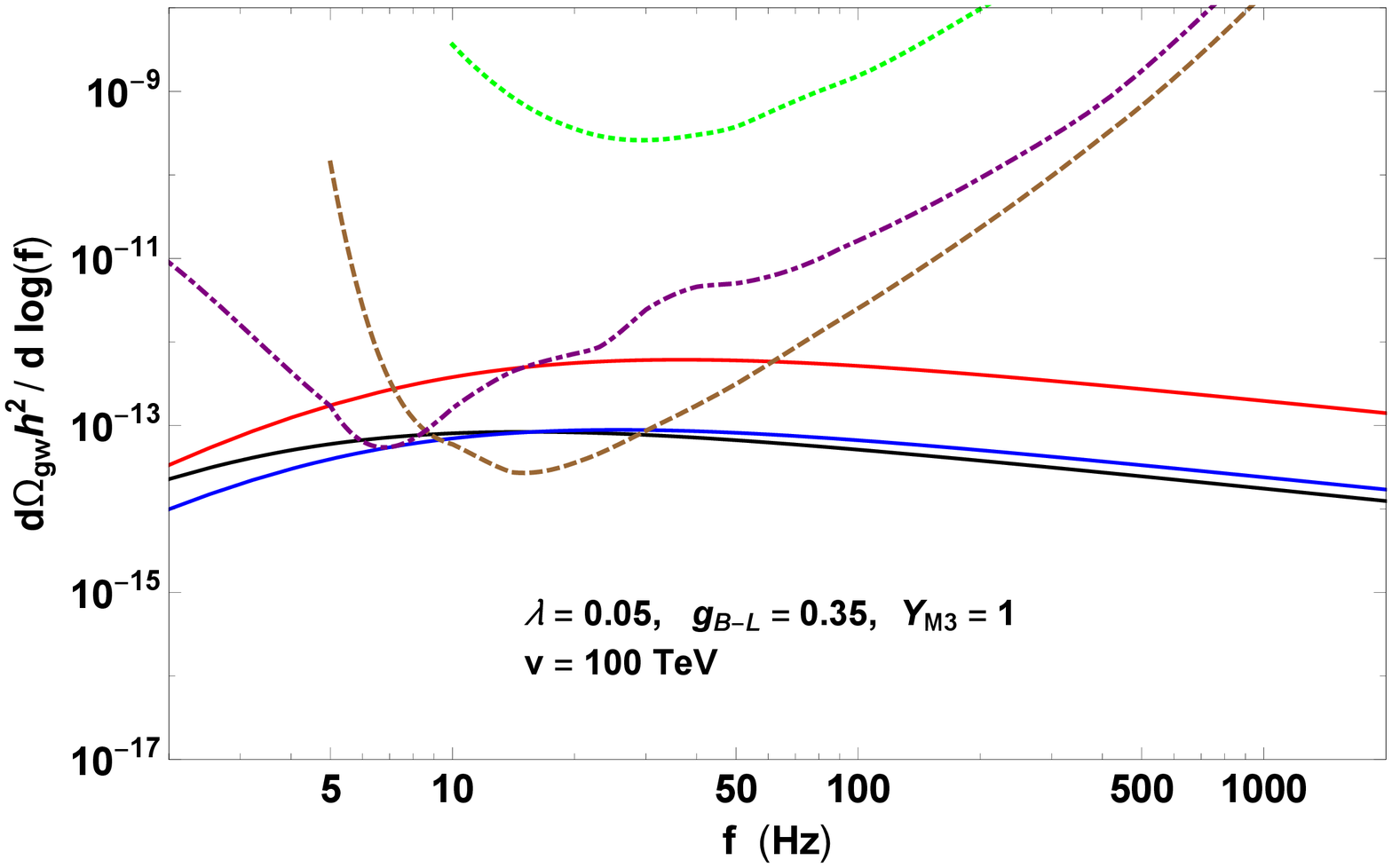}
    \\
    \includegraphics[width=80mm]{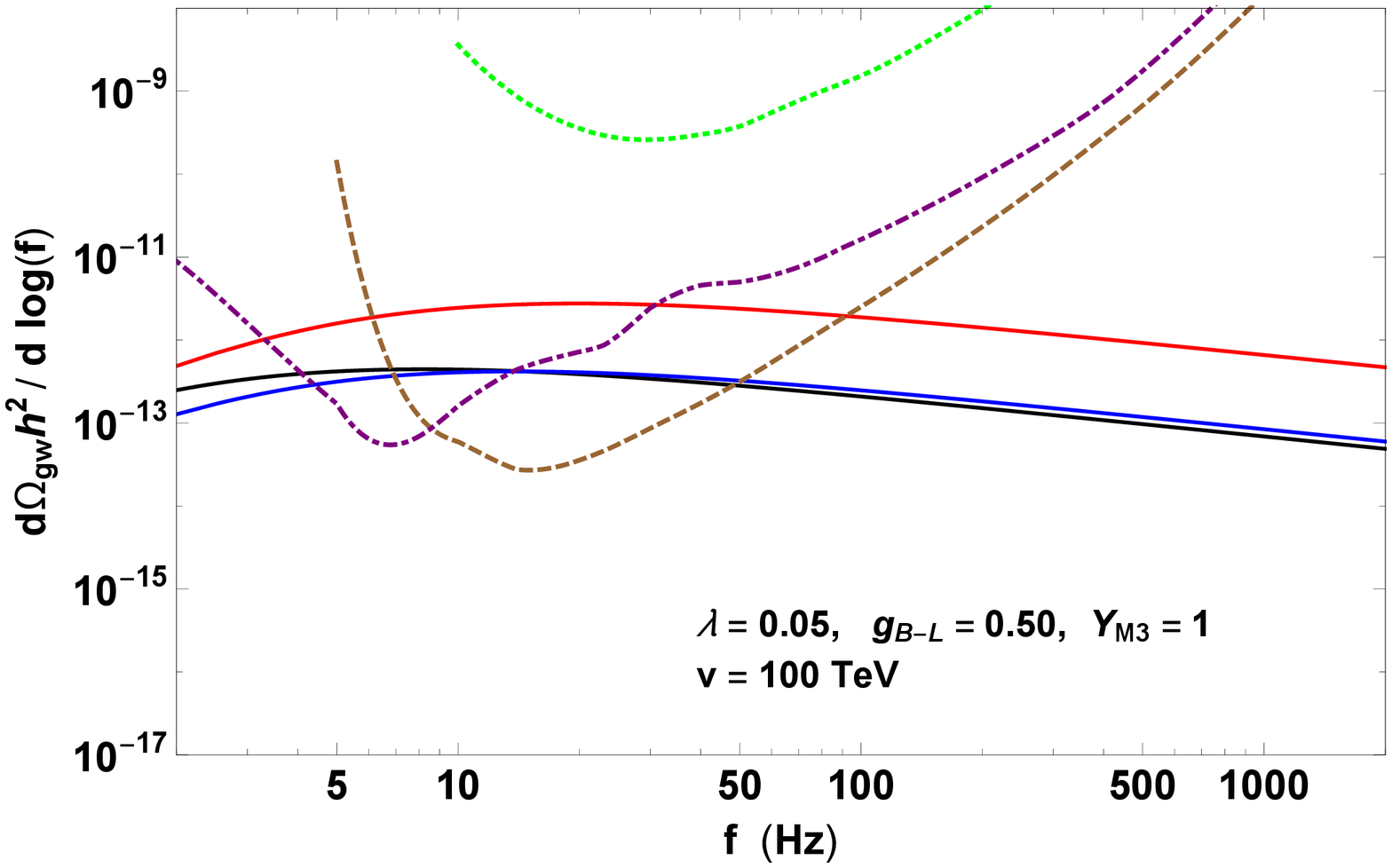}
    \includegraphics[width=80mm]{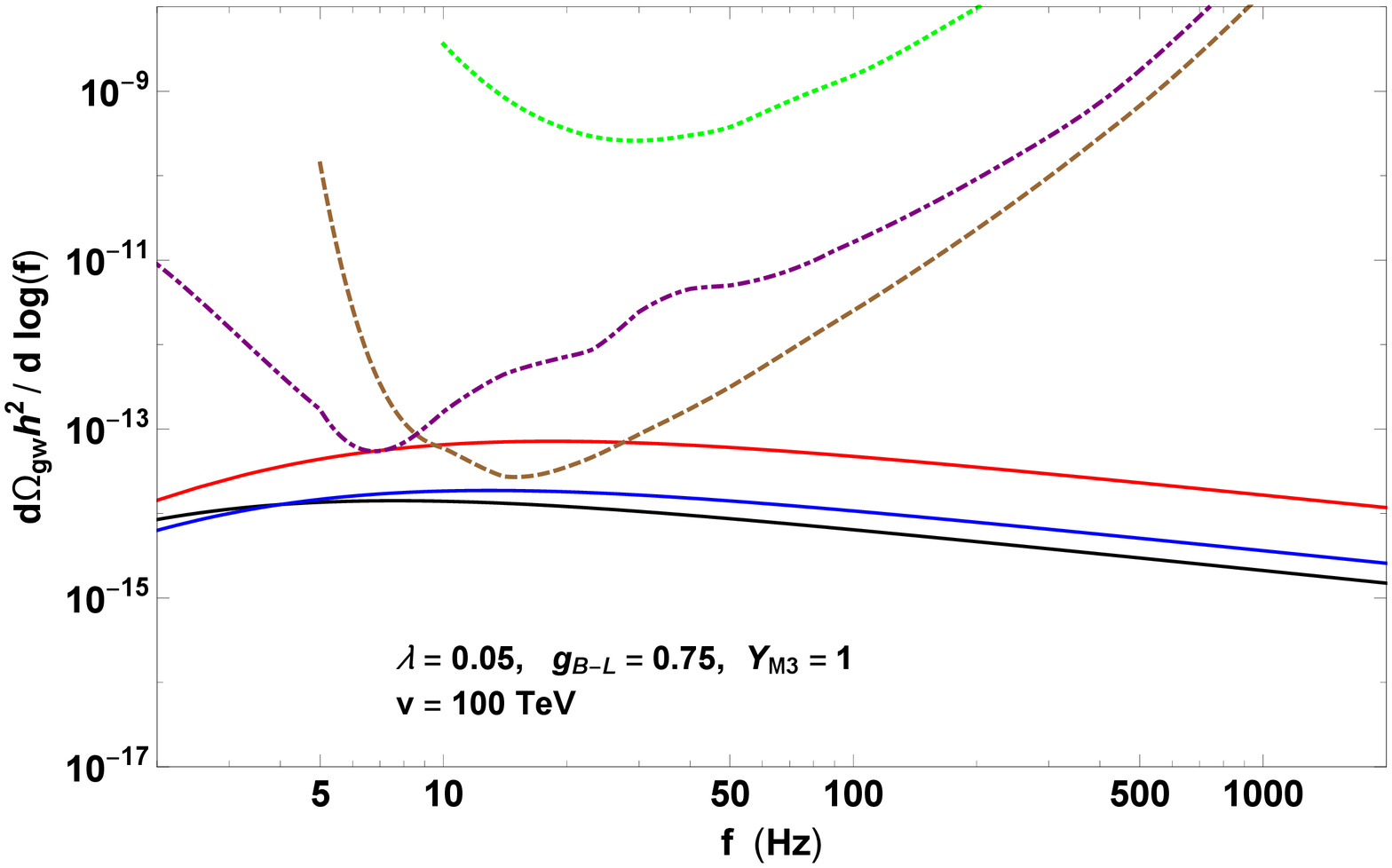}
    \caption{
    Energy spectrum of stochastic gravitational waves from a $U(1)_{B-L}$ breaking phase transition in the case when $\lambda=0.05$
    and $v=100$~TeV.
    From the upper-left to lower-right, each plot corresponds to different values of the $U(1)_{B-L}$ gauge coupling constant, 
    $g_{B-L}=0.2,~0.35,~0.5,~0.75$.
    We fix $Y_{M3}=1$ and $Y_{M1}=Y_{M2}=0$.
    In each plot, the black-solid, red-solid and blue-solid lines correspond to different assumptions on the bubble wall velocity
    with $v_{\rm w}=0.92,~0.72,~0.44$, respectively.
    The green-dotted, purple-dot-dashed and brown-dashed lines respectively represent
     the Advanced LIGO design sensitivity and the sensitivities of Einstein Telescope (ET-D estimate) and Cosmic Explorer
     for frequency bin of 0.25~Hz and 2~years of data collection.
    We note that
     the spectrum approximately slides with $v$, with the peak position proportional to $v$ 
     and the strength and shape unaltered.
    }
    \label{gw}
  \end{center}
\end{figure}

Our estimate on the total gravitational wave spectrum is presented in Fig.~\ref{gw}, for $\lambda=0.05$, $v=100$~TeV,
 $Y_{M3}=1$ and $Y_{M1}=Y_{M2}=0$, and for various values of $g_{B-L}$.
The spectrum is given in terms of frequency $f=k/(2\pi)$.
The design sensitivity of Advanced LIGO and the sensitivities of Einstein Telescope and Cosmic Explorer 
 for frequency bin of $\delta f=0.25$~Hz and ${\cal T}=$2~years of data collection,
 are estimated from Refs.~\cite{Martynov:2016fzi,Hild:2010id,Evans:2016mbw}
 through the relation d$\Omega_{\rm gw}(f)/$d$\log f=(1/\sqrt{2\delta f {\cal T}})2\pi^2f^3S_h(f)/(3H_0^2)$,
 where $S_h(f)$ denotes strain power spectral density and $H_0$ denotes the Hubble rate today.
These sensitivity curves are overlaid on the plots.
The spectrum around the peak, which is relevant to gravitational wave detection, slides with $v$,
 with the peak position proportional to $v$ and the strength unaltered.
This is because $T_{\rm n}$ is proportional to $v$ and $\alpha_\theta(T_{\rm n}),\,\beta_{\rm n}/H_{\rm n}$ are independent of $v$
 when we fix $\mu=v$ and fix the right-hand side of Eq.~(\ref{tnestimate}).
Although the contribution of turbulence~Eq.~(\ref{turb}) depends on $v/M_*$, it is negligible around the peak. 
Therefore, the spectrum around the peak depends on $v$ only through the combination $f/v$.

Lowering $Y_{M3}$ reduces the strength of the spectrum, but does not significantly change its shape and position.
This is because $\alpha_\theta(T_{\rm n})$ decreases for $Y_{M3}=0.1$, while $T_{\rm n}$ and $\beta_{\rm n}/H_{\rm n}$
 have little or no dependence on $Y_{M3}$ when $g_{B-L}\geq0.2$, as seen in the right panels of Figs.~\ref{TnPlot},\ref{alphaPlot},

We find that in our benchmark with $\lambda=0.05$,
 stochastic gravitational waves are out of reach of the Advanced LIGO design sensitivity for all values of $g_{B-L}$.
However, for values of the $U(1)_{B-L}$ gauge coupling constant near the weak gauge coupling constant, such as $g_{B-L}=0.5$,
 stochastic gravitational waves can be detected at future Einstein Telescope and Cosmic Explorer.
Noting that the spectrum around the peak slides with the $U(1)_{B-L}$-breaking VEV $v$,
 we see that Einstein Telescope and Cosmic Explorer cover a wide range of the $U(1)_{B-L}$-breaking VEV, 
 which is estimated to be $v\lesssim1000$~TeV.

We can utilize the position and strength of the peak of the gravitational wave spectrum, to relate 
 the $U(1)_{B-L}$ gauge coupling constant $g_{B-L}$ with the $U(1)_{B-L}$-breaking VEV $v$. It proceeds as follows:
For a fixed value of $g_{B-L}$, 
  $\alpha_\theta(T_{\rm n})$ has violent dependence on $\lambda$,
  while it has much milder dependence on the Majorana Yukawa coupling $Y_{Mj}$ (see Figs.~\ref{alphaPlot},\ref{lambdadep}).
Also, $\beta_{\rm n}/H_{\rm n}$ depends only weakly on $\lambda$ (see Fig.~\ref{lambdadep}).
Therefore, we can estimate the superpotential coupling $\lambda$ from the strength of the gravitational wave spectrum at the peak
 through the $\lambda$-dependence of $\alpha_\theta(T_{\rm n})$
 (see Eq.~(\ref{sw}) and note that the spectrum around the peak is dominated by sound waves' contribution).
Once $g_{B-L}$ and $\lambda$ are known, we can determine $T_{\rm n}/v$ (along with $\beta_{\rm n}/H_{\rm n}$),
 thereby relating the peak position to the $U(1)_{B-L}$-breaking VEV $v$.
The above correspondence between $g_{B-L}$ and $v$ obtained from the gravitational wave spectrum
 offers a clue about the minimal SUSY $U(1)_{B-L}$ model, complementing future collider searches for the $U(1)_{B-L}$ gauge boson.
\\

%%%%%%%%%%%%%%%
\section{Summary}

We have studied the phase transition of a $U(1)$ gauge symmetry breaking in a SUSY model and the production of stochastic gravitational waves associated with it.
We have concentrated on a particular model, which is the minimal SUSY $U(1)_{B-L}$ model with $R$-symmetric superpotential.
We have worked in the SUSY limit by assuming that the nucleation temperature is above SUSY breaking scale so that soft SUSY breaking terms are negligible.
We have derived the finite temperature effective potential for the $U(1)_{B-L}$-breaking VEVs $h,\bar{h}$,
 and computed the $O(3)$-symmetric Euclidean action of a high-temperature $U(1)_{B-L}$-breaking multi-field phase transition.
We have estimated stochastic gravitational waves generated from the phase transition 
 in the case with $\lambda=0.05$, where $\alpha_\theta(T_{\rm n})<0.1$ and well-established study on gravitational wave production
 is available.
We have found that for values of the $U(1)_{B-L}$ gauge coupling constant around $g_{B-L}\simeq0.5$,
 and for a wide range of the $U(1)_{B-L}$-breaking VEV $v\lesssim1000$~TeV,
 stochastic gravitational waves can be detected at future Einstein Telescope and Cosmic Explorer.
We point out that the position and strength of the peak of the gravitational wave spectrum provides information on the relation between $g_{B-L}$ and $v$.
\\

%%%%%%%%%%%%%%%
\section*{Acknowledgement}
This work is partially supported by Scientific Grants by the Ministry of Education, Culture, Sports, Science and Technology of Japan,
Nos.~17K05415, 18H04590 and 19H051061 (NH), and No.~19K147101 (TY).
\\

%%%%%%%%%%%%%%%
%%% References %%%
%%%%%%%%%%%%%%%

\end{document}